\documentclass[12pt,preprint]{aastex}

\usepackage{booktabs}
\usepackage{flushend} 
\newcommand{\VLSR}{V_{\rm LSR}}
\newcommand{\FeII}{[\ion{Fe}{2}]}

\newcommand{\kms}{km~s$^{-1}$}

\newcommand{\degree}{^{\circ}}
\newcommand{\Av}{A_{\rm V}}
\newcommand{\Ak}{A_{\rm K}}

\newcommand{\myemail}{hyoh@astro.snu.ac.kr}

\shorttitle{IGRINS Datacube on Orion KL}
\shortauthors{Oh et al.}

\begin{document}


\title{Three-dimensional Shock Structure of Orion KL Outflow with IGRINS\footnote{This paper includes data taken at The McDonald Observatory of The University of Texas at Austin.}}


\author{Heeyoung Oh\altaffilmark{1,2,3}, Tae-Soo Pyo\altaffilmark{4,5}, Kyle Kaplan\altaffilmark{6}, In-Soo Yuk\altaffilmark{1}, Byeong-Gon Park\altaffilmark{1,2}, Gregory Mace\altaffilmark{6}, Chan Park\altaffilmark{1}, Moo-Young Chun\altaffilmark{1}, Soojong Pak\altaffilmark{7}, Kang-Min Kim\altaffilmark{1}, Jae Sok Oh\altaffilmark{1}, Ueejeong Jeong\altaffilmark{1}, Young Sam Yu\altaffilmark{1},
Jae-Joon Lee\altaffilmark{1}, Hwihyun Kim\altaffilmark{1,6}, Narae Hwang\altaffilmark{1}, Hye-In Lee\altaffilmark{7}, Huynh Anh Nguyen Le\altaffilmark{3,7}, Sungho Lee\altaffilmark{1}}
\and
\author{Daniel T. Jaffe\altaffilmark{6}}
 
%
%


\altaffiltext{1}{Korea Astronomy and Space Science Institute, 776 Daedeok-daero, Yuseong-gu, Daejeon 34055, Korea. {\myemail}}
\altaffiltext{2}{Korea University of Science and Technology, 217 Gajeong-ro, Yuseong-gu, Daejeon 34113, Korea.}
\altaffiltext{3}{Department of Physics and Astronomy, Seoul National University, 1 Gwanak-ro, Gwanak-gu, Seoul 08826, Korea.}
\altaffiltext{4}{Subaru Telescope, National Astronomical Observatory of Japan, National Institutes of Natural Sciences (NINS), 650 North A'ohoku Place, Hilo, HI 96720.}
\altaffiltext{5}{School of Mathematical and Physical Science, SOKENDAI (The Graduate University for Advanced Studies), Hayama, Kanagawa 240-0193, Japan.}
\altaffiltext{6}{Department of Astronomy, University of Texas at Austin, Austin, TX 78712.}
\altaffiltext{7}{School of Space Research and Institute of Natural Sciences, Kyung Hee University, 1732 Deogyeong-daero, Giheung-gu, Yongin-si, Gyeonggi-do 17104, Korea.}


\begin{abstract}
We report a study of the three-dimensional (3D) outflow structure of a 15$\arcsec$ $\times$ 13$\arcsec$ area around H$_{2}$ peak 1 in Orion KL with slit-scan observations (13 slits) using the Immersion Grating Infrared Spectrograph.
The datacubes, with high velocity-resolution ($\sim$ 7.5 {\kms}) provide high contrast imaging within ultra-narrow bands, and enable the detection of the main stream of the previously reported H$_{2}$ outflow fingers.
We identified 31 distinct fingers in H$_{2}$ 1$-$0 S(1) $\lambda$2.122 $\micron$ emission. The line profile at each finger shows multiple-velocity peaks with a strong low-velocity component around the systemic velocity at ${\VLSR}$ =  $+$8 {\kms} and high velocity emission ($|$${\VLSR}$$|$ =  45$-$135 {\kms}) indicating a typical bow-shock. The observed radial velocity gradients of $\sim$ 4 {\kms} arcsec$^{-1}$ agree well with the velocities inferred from large-scale proper motions, where the projected motion is proportional to distance from a common origin. We construct a conceptual 3D map of the fingers with the estimated inclination angles of 57$\degree$$-$74$\degree$. The extinction difference ($\Delta$$A_{\rm v}$ $>$ 10 mag) between blueshifted and redshifted fingers indicates high internal extinction. The extinction, the overall angular spread and scale of the flow argue for an ambient medium with very high density (10$^{5}$$-$10$^{6}$ cm$^{-3}$), consistent with molecular line observations of the OMC core.
The radial velocity gradients and the 3D distributions of the fingers together support the hypothesis of simultaneous, radial explosion of the Orion KL outflow.
\end{abstract}





\section{Introduction}\label{sec:ori_intro}

The Becklin-Neugebauer/Kleinmann-Low \citep[BN/KL,][]{Becklin1967,Kleinmann1967} nebula is a spectacular southeast-northwest outflow region that lies behind the Orion nebula. The distance is 414 $\pm$ 7 pc \citep{Menten2007}.
The outflow activity was revealed in molecular emission lines such as CO \citep{Kwan1976}, HCO$^{\rm +}$, SO, HCN, and SiO \citep[e.g.,][]{Welch1981,Knapp1981,Plambeck1982}. Their line profiles showed very broad line width (full width zero intensity (FWZI) $\sim$ 190 {\kms}). The detections of various maser emission lines of H$_{2}$O, OH, SiO, and CH$_{3}$OH and evidence from proper motion measurements of some of these lines also indicated the presence of strong outflow in this region \citep{Genzel1981,Cohen2006,Matthews2010,Peng2012}. Herbig-Haro (HH) objects \citep{Axon1984}, and finger-like molecular hydrogen (H$_{2}$) and forbidden iron ({\FeII}) features were discovered at optical and near-IR wavelengths. \citep{Taylor1984,Allen1993}. \citet{Bally2011} showed more than 100 individual fingers from the high spatial resolution imaging.

The driving mechanism of this powerful system remains unclear despite intensive study. \citet{Bally2005} and \citet{Bally2011} suggested that the Orion BN/KL outflows are powered by the dynamical decay of a non-hierarchical multiple system, as evinced by the radio proper motions of some of the massive young stellar objects (YSOs) in this region: the BN object, radio sources I, and n \citep{Menten1995,Gomez2008}.
\citet{Bally2015} indicated the position of ejection center lies within 1$\arcsec$ of J2000 $=$ 05:35:14.360, $-$05:22:28.70, which was originally derived from the intersection of the radio proper motions of three sources above \citep{Gomez2005,Gomez2008}.
\citet{Tan2004} and \citet{Chatterjee2012} argued for a scenario in which the BN object was ejected from the Trapezium about 4000 years ago, and the passed through the Orion molecular cloud (OMC-1) hot core and source I about 500 years ago. The dynamical age of the outflow ranges from 500 to larger than 1000 years, based on proper motion observations in optical, near-IR, and millimeter waveband \citep{Doi2002,Gomez2008,Bally2011,Wu2014}. 
The proper motions of fingers measured in the near-IR and the optical show that the velocities of fingers are proportional to the distance from the ejection center, implying that they originated from a single explosive event \citep{Doi2002,Bally2011}.

H$_{2}$ emission is a prominent feature in the Orion KL outflow. H$_{2}$ is an effective coolant in shocks and is a useful tool for studying both the kinematics of and shock conditions in molecular outflows. The {\FeII} emission, mainly detected in ``fingertips'', usually traces higher velocities than H$_{2}$ flows \citep{Pyo2002,Davis2003}. The study of near-IR emission of bow shocks shows that C-type shocks produce H$_{2}$ emission in the bow wings, while dissociative J-type shocks produce {\FeII} emission near the bow apex \citep{O'Connell2005}.

The BN/KL region is very complicated with many overlapping outflow features. Spectroscopic studies combined with spatial information allow us to learn about the overall structure of the outflows and  provides a key to understand the formation mechanism for the outflows.
There have been spectroscopic maps of H$_{2}$ emission with mid- to high-spectral resolutions using a Fabry-Perot (FP) interferometer in OMC-1 \citep{Usuda1996,Chrysostomou1997,Salas1999}, which interpreted the structure of velocity distribution, line profiles, and shock excitation.
\citet{Tedds1999} reported the high spatial resolution long-slit spectroscopy of H$_{2}$ and {\FeII} lines using CGS4/UKIRT towards two selected bullets of M42 HH126-053 and M42 HH120-114.
\citet{Youngblood2016} reported the more complete near-IR position-position-velocity cubes with slit-scan observations those cover 2$\farcm$7 $\times$ 3$\farcm$3 area on the Orion BN/KL outflow. With a spectral resolution of $\lambda$/$\Delta$$\lambda$ $\sim$ 3500, they found velocity structure consistent with a 500-year-old outflow.

In this study, we report the results of spectroscopic mapping made with higher spectral resolution than in any previous study. Using the Immersion GRating INfrared Spectrograph (IGRINS), we obtained a map of the region around OMC-1 H$_{2}$ peak 1 \citep{Beckwith1978} with consecutive multiple-slit positions.
IGRINS has velocity resolution ($\Delta v$) about 7.5 {\kms}. In addition to high spectral resolution, IGRINS provides greater dynamic range through superb surface brightness sensitivity and a clean spectral point spread function. FP interferometers provide high resolution but strong wing of their Lorentzian line profiles make it hard to detect weak high velocity emission in the presence of strong central emission. We detected the main stream of the fingers as many narrow, linear patterns in the channel maps. The conventional narrow-band filter images usually show only the boundary shape of bow fingers, because the filter width corresponds to several thousands {\kms} in velocity. Our channel maps of high spectral-resolution provides high contrast imaging with ultra-narrow band width ($\Delta$$v$ $\sim$ 10 {\kms}). We identify 31 distinct outflow fingers around the peak 1 region that are spatially overlapped but resolved in the datacubes of H$_{2}$ lines.
We analyze the physics of individual fingers by comparing with the high angular resolution image taken with GSAOI at Gemini South \citep{Bally2015}.
We generate three-dimensional (3D) pictures of subregions in combination with estimated inclination angles ($i$) of outflows. 


\section{Observation and data reduction} \label{sec:ori_obs}

The data were obtained on 2014 December 1 (UT) with IGRINS \citep{Yuk2010,Park2014} mounted on the 2.7m Harlan J. Smith Telescope at the McDonald Observatory of the University of Texas at Austin. IGRINS is a cross-dispersed near-IR spectrograph using a silicon immersion echelle grating. The whole wavelength range of the infrared H- and K-bands (1.49 $-$ 2.46 $\micron$) are observed simultaneously, with a spectral resolving power $R$ $\equiv$ $\lambda/\Delta\lambda$ $\sim$ 45,000. 
The slit size was 1\farcs0 $\times$ 15\farcs0. The resolving power corresponds to a velocity resolution ($\Delta v$) of 7.5 {\kms}, with $\sim$ 3.5 pixel sampling. The pixel scale is 0\farcs24 $-$ 0\farcs29 pixel$^{-1}$ along the slit, the value is larger in higher orders.  Auto-guiding was performed during each exposure with a K-band slit-viewing camera (pixel scale = 0\farcs12 pixel$^{-1}$). The guiding uncertainty was smaller than 0\farcs4 on average. The K-band seeing during the observations was $\sim$ 0\farcs9. 

\begin{figure}
\epsscale{.60}
\plotone{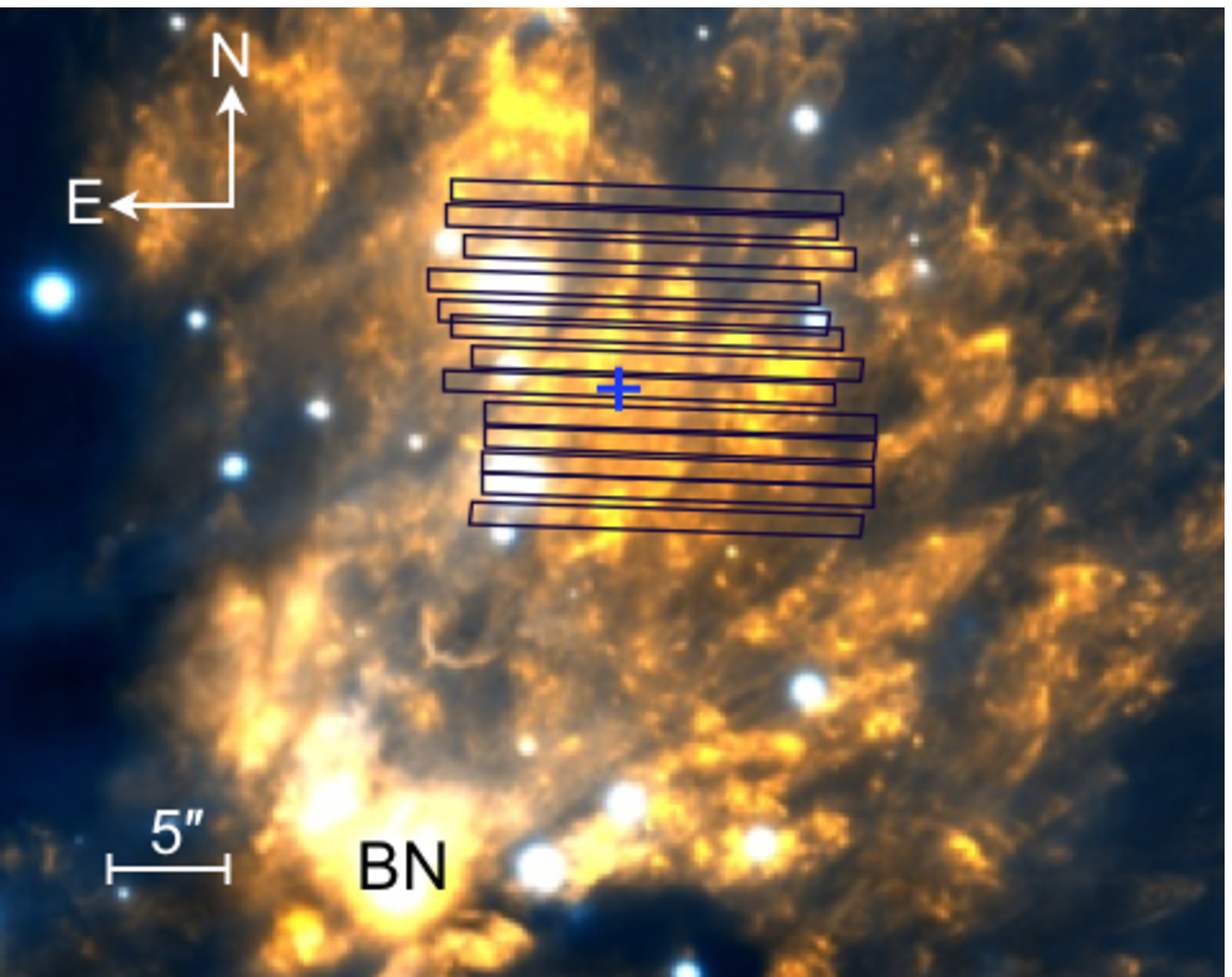}
\caption[IGRINS slit positions.]{IGRINS slit positions. Data were obtained at 13 different positions. The blue cross is the position of H$_{2}$ peak 1 in OMC-1 \citep{Sugai1994}. Slit size is 1\farcs0 (W) $\times$ 15\farcs0 (L). The background image is a color composite image with H$_{2}$ 1$-$0 S(1) (Orange) and {\FeII} (Blue) emission lines \citep{Bally2015}. \label{fig:slitposition}}
\end{figure}

By performing a slit-scanning observation at 13-slit positions with $\sim$ 1$\arcsec$ step perpendicular to the slit length, we covered $\sim$ 15$\arcsec$ $\times$ 13$\arcsec$ area including H$_{2}$ peak 1 \citep[J2000 $=$ 05:35:13.57, $-$05:22:03.8,][]{Sugai1994} in OMC-1. The slit positions on the sky are  shown in Figure \ref{fig:slitposition}. The slit position angle (P.A.) was 88$\arcdeg$ for all slit positions, which was confirmed by comparing slit-view images to 2MASS K-band images. On-source exposure time was 300 s at each slit position. Off-source frames with same exposure times were obtained between every third on-source observations, at a position of 1800$\arcsec$ south and 1800$\arcsec$ west of peak 1. HR 1724, which has K magnitude of 6.30 and spectral type of A0V, was observed as a telluric standard star. We took Th-Ar and halogen lamp frames for wavelength calibration and flat-fielding, respectively.

The basic data reduction was done using the IGRINS Pipeline Package\footnote{The IGRINS Pipeline Package is downloadable at \url{https://github.com/igrins/plp}. (doi:10.5281/zenodo.18579).} (PLP). The PLP performs sky subtraction, flat-fielding, bad pixel correction, aperture extraction, and wavelength calibration. For the processing of two-dimensional (2D) spectra from IGRINS data, a software called Plotspec\footnote{https://github.com/kfkaplan/plotspec} has been developed. Plotspec provides continuous 2D spectra of all the IGRINS H- and K-band orders, removal of stellar photospheric absorption lines from the standard star, telluric correction, and relative flux calibration, etc. Continuum is subtracted using the pixel values obtained by a robust median filter running in wavelength direction.
With the Plotspec code, we also construct a 3D datacube from the slit-scan data. 
The gaps between slits are filled with the median pixel values from the adjacent point.
We sampled every $\sim$ 1$\arcsec$ along the direction slit width.
The angular resolution along the slit length is seeing limited.
We used the FLUXER tool\footnote{Interactive IDL routine written by Christof Iserlohe, http://www.ciserlohe.de/fluxer/fluxer.html} to extract position-velocity diagrams (PVDs) at a desired slit P.A. from the datacube.




\section{Results} \label{sec:ori_result}
In the 1.49$-$2.46 $\micron$ range, we detected more than 30 H$_{2}$ lines (Table \ref{tbl:H2flux}) and eight {\FeII} lines arising in the Orion KL outflow. In this section, we report the analysis of the datacube constructed from every detected line. We examine the characteristics of the bow-shape ``bullets'' using channel maps, PVDs, and line profiles. The hydrogen population diagram extracted from the datacubes also allow a study of the shock properties at distinct space-velocity positions. Since {\FeII} lines are very weak, except for the $a^{4}D_{7/2} - a^{4}F_{9/2}$ $\lambda$1.644 $\micron$ line, we use this line to compare to H$_{2}$ lines.

\subsection{Molecular hydrogen lines} \label{h210s1}

\subsubsection{Channel maps and identification of finger structures} \label{sec:ori_channelmap}
Figure \ref{fig:arrows} shows an H$_{2}$ 1$-$0 S(1) emission image integrated over a velocity range of $-$150 {\kms} $<$ ${\VLSR}$ $<$ $+$150 {\kms} and the slit-scan area overlaid on a high-resolution H$_{2}$ 1$-$0 S(1) emission image taken with Gemini GSAOI \citep{Bally2015}. Overall, the integrated intensity distribution is well matched with the high-resolution image, which shows many small bow features in the Orion KL peak 1 area. The presence of unsubtracted stellar continuum in the GSAOI image accounts for much of the difference. Figure \ref{fig:10s1channel} displays the H$_{2}$ 1$-$0 S(1) line channel maps at 10 {\kms} intervals for $-$140 {\kms} $<$ ${\VLSR}$ $<$ $+$100 {\kms}.

We constructed a datacube with 1$\arcsec$$\times$1$\arcsec$$\times$1 {\kms} pixels.
At low velocities, the large amount of spatial overlap leaves the picture quite confused. At high velocities, however, by comparing adjacent velocity channels in the datacube, we detected several tens of narrow, linear features with strong velocity gradients, stretched along the southeast-northwest direction at $|$$\VLSR$$|$ $>$ 40 {\kms}. They have lengths of 2$\arcsec$$-$4$\arcsec$ in the plane of the sky, and are distinct in velocity and space. Their line widths are narrow, usually $<$ 30 {\kms}. We infer that these are velocity-resolved main streams of fingers which are spatially overlapped along the line of the sight (see also Section \ref{sec:ori_3D}).
We identified 24 blueshifted and 7 redshifted outflow fingers with following criteria: 1) They show clumpy features continuously stretched over more than 2$\arcsec$, 2) they have intensity level above 2$\sigma$, and 3) each finger should have its own peak velocity.

The identified streams are marked in Figures \ref{fig:arrows} and \ref{fig:10s1channel}.
We marked the blueshifted and redshifted streams with blue and red lines, respectively. The size of the lines represent the apparent lengths of identified fingers. For the fingers located at the boundary of the slit-scan area, we measured their lengths for the portions covered within the field. Most identified streams are coincident with the locations of the small bow fingers in the high-resolution image. In Figure \ref{fig:ballyvector}, we superposed identified streams on Figure 3 of \citet{Bally2015}. It shows that the directions of the outflow streams are almost parallel to the large-scale vectors that connect the ejection center and the outermost fingers. The finger identification numbers (FIDs) marked on figures increase with radial velocity from $-$127 to $+$88 {\kms}.
Several fingers form linear features in space and velocity. FIDs 20$-$23 and FIDS 9, 11, 12 are such cases. These can be considered as groups forming a larger finger structure, while each of them shows a distinct peak velocity.  
We note that ``finger'' in this study indicates a short linear structure, which is different from its traditional definition, where the larger scale ``fingers'' each contain multiple bows.
We could not resolve the finger (stream) patterns at $-$30 {\kms} $<$ ${\VLSR}$ $<$ $+$30 {\kms} due to the strong amorphous diffuse emission (see Figure \ref{fig:10s1channel}). The shape of this low-velocity emission changes dynamically over the channel maps. This emission likely comes from blended bows that are either slower moving or in the plane of the sky. They are fully blended spatially at our resolution. For reasons not fully understood, we note that the boundary of this diffuse emission forms a ring shape in the channel maps centered at $-$5 and $+$5 {\kms}.

\subsubsection{Position velocity diagrams (PVDs) and line profiles} \label{pvd}
We extracted the PVDs from the datacube using the FLUXER tool (see also Section \ref{sec:ori_obs}). As shown in Figure \ref{fig:ballyvector}, we choose 9 different directions along directions showing large-scale vectors. The pixel values are interpolated along each given slit direction. In the extracted PVDs, we use linear interpolation to increase the number of pixels on both the velocity and space axes by a factor of 5.
The PVDs of the H$_{2}$ 1$-$0 S(1) emission line are shown in Figure \ref{fig:pvd}. Low-velocity emission, at $-$40 {\kms} $<$ ${\VLSR}$ $<$ $+$30 {\kms}, always forms the majority of the H$_{2}$ flux at any given position.
For the high-velocity components, we compared and matched their positions and velocities to the fingers we identified in the datacube. The white dashed lines and the numbers mark fingers and FIDs found in PVDs. The velocity gradients of high-velocity components in Figure \ref{fig:pvd} are similar in all fingers, with the absolute velocity decreasing at a rate of 2$-$6 {\kms} arcsec$^{-1}$ from fingertip toward the ejection center.

In Figure \ref{fig:10s1profile}, we show H$_{2}$ 1$-$0 S(1) emission line profile at each fingertip for FIDs from 1 to 31.
Every fingertip shows multiple peaks: strong low-velocity peak around the systemic velocity of $\VLSR$ $=$ $+$8 {\kms} \citep{Chrysostomou1997} and peaks at higher and lower velocities ($|$$\VLSR$$|$ = 45 $-$ 135 {\kms}). Several line profiles show three peaks due to overlap of another finger in the sampled region. The high-velocity peaks are marked with solid vertical lines at each panel in Figure \ref{fig:10s1profile}. From many previous studies, including \citet{Bally2015}, we know that there are more than 100 bow-shock bullets and multiple-peak profiles in this region \citep[e.g.,][]{Salas1999}. In Table \ref{tbl:fingerlist}, we listed the FIDs and peak velocities of the high-velocity components. We estimated FWZI of double-peak line profiles. In Section \ref{sec:ori_discussion}, we discuss detailed kinematics of each bow fingers, especially the velocity gradient of the high-velocity component along the outflow direction. The 3D distribution of outflow streams will also be discussed.

\subsubsection{Extinction}\label{sec:ori_extinction}
We estimated the extinction toward peak 1 using the ratios between pairs of H$_{2}$ lines that arise from the same upper level. Three pairs, v $=$ 1-0: Q(3) $\lambda$2.424 $\micron$ / S(1) $\lambda$2.122 $\micron$, Q(2) $\lambda$2.413 $\micron$ / S(0) $\lambda$2.223 $\micron$, and Q(4) $\lambda$2.437 $\micron$ / S(2) $\lambda$2.034 $\micron$ are used. The transition probabilities are taken from \citet{Turner1977}. We adopt extinction law $A_{\rm \lambda}$ = $\Av$(0.55$\micron$/$\lambda$)$^{1.6}$ \citep{Rieke1985}. Figure \ref{fig:Av}b shows the visual extinction ($\Av$) estimated from the intensity ratios between monochromatic images of the emission lines over velocity range of  $\pm$150 {\kms}. The median value from the three line ratios is used in the estimation. It shows spatial distribution in the range of 0 $<$ $\Av$ $<$ 8, which corresponds to $\Ak$ = 0$-$1. This is similar to the value obtained in previous studies \citep[e.g.,][]{Rosenthal2000,Youngblood2016}. However, we found that the $\Av$ values estimated in different velocity ranges show a large deviation. Figure \ref{fig:Av}c shows that $\Av$ varies over velocity channel maps in the range of 0 $<$ $\Av$ $<$ 15 mag.
We listed measured $\Av$ at the position of every bow finger in Table \ref{tbl:fingerlist}. We found relatively small $\Av$ values ($\sim$ 0) at the highest blueshifted velocity and large $\Av$ values ($\Av$ $=$ 7.4$-$15.1 mag) at redshifted $\VLSR$ velocities ($\VLSR$ $>$ $+$65 {\kms}). This difference implies a differential extinction along the line of sight. 
To investigate the $\Av$ along the line of sight, we need to consider not only the radial velocity shown in channel maps, but also the inclination angles of the fingers. In Section \ref{sec:ori_Av}, we discuss this more in connection with the relative depth of the fingers.

\subsubsection{Shock condition and population diagram} \label{sec:ori_ratio}
Figure \ref{fig:ratiochannel} shows the channel maps of H$_{2}$ 2$-$1 S(1) / 1$-$0 S(1) ratio. This ratio is a commonly used indicator to distinguish the excitation mechanism, where the typical ratios for excitation by shocks and UV radiation are 0.05$-$0.27 and 0.55, respectively \citep{Smith1995,Black1987,Pak1998}. In each channel, the reddening was corrected using the $\Av$ maps shown in Figure \ref{fig:Av}c. The line ratio agrees well with the shock excited case, ranging from 0.05 to 0.14. The ratio around the systemic velocity ($\VLSR$ = $+$8 {\kms}) is slightly higher than the ratio expected from a pure C-type shock, which is $\sim$ 0.05 \citep{Smith1995}. Some high-velocity components show higher line ratios, e.g., $\sim$ 0.15 and $\sim$ 0.12 at $-$135 and $+$65 {\kms} respectively. This indicates a mixture of C- and J-type shock components, where the pure J-shock ratio is $\sim$ 0.27 \citep{Smith1995}.

An H$_{2}$ state population diagram constructed from the various H$_{2}$ emission lines allows us to study the rotational  and vibrational state of the gas \citep{Black1976,Beckwith1978}. In Figure \ref{fig:cdr}, we show population diagrams deduced from 7 velocity ranges and positions, which are marked with green boxes in H$_{2}$ 2$-$1 S(1) / 1$-$0 S(1) ratio channel maps in Figure \ref{fig:ratiochannel}. The selected positions are the locations of bright shock emission in chosen velocity channel maps (Figure \ref{fig:10s1channel}).
From datacubes of 35 detected H$_{2}$ lines, we extracted the relative intensities at selected pixel areas and at velocity ranges of $V$$_{\rm central}$ $\pm$ 5 {\kms}. In the plot, the column densities are normalized to those derived for the H$_{2}$ 1$-$0 S(1) line and are relative to the Boltzmann distribution at 2,000K. The line intensities are reddening corrected with $A_{\rm \lambda}$ at the same velocity channels of Figure \ref{fig:Av}c. $\Ak$ is indicated in each panel in Figure \ref{fig:cdr}. In the plot, we excluded the lines with a low signal-to-noise ratio (S/N$<$2) and the lines affected by OH sky emission or telluric absorptions.

The population diagrams indicate thermalization with population trends following a single line. The excitation temperature can be derived from the slope of the populations versus upper state energy. At $\VLSR$ $=$ $+$5 {\kms}, close to the systemic velocity ($+$8 {\kms}), the rotational temperature ($T_{\rm rot}$) is 1800 (v = 1), 2600 (v = 2), and 3200 K (v = 3), where v is the upper vibrational level of transition lines. In the high-velocity regions, $T_{\rm rot}$ at v = 1 is similar to that at the systemic velocity, while $T_{\rm rot}$ at v = 2 and 3 show various values among 2000$-$3000 K with larger uncertainty ($>$ 400 K). The estimated $T_{\rm rot}$ are similar to those from other shocked outflows in low- or intermediate-mass star formation region \citep{Nisini2002,Takami2006,Oh2016}.
In Section \ref{sec:ori_discussion}, we discuss the further interpretation of line ratios in relation to the various shock models.

\subsection{{\FeII} $\lambda$1.644 $\micron$ emission line} \label{sec:ori_feii}
Figure \ref{fig:Fechannel} shows channel maps for the {\FeII} $\lambda$1.644 $\micron$ emission line, in the same velocity intervals as for H$_{2}$ 1$-$0 S(1) in Figure \ref{fig:10s1channel}. The relative intensity of {\FeII} $\lambda$1.644 $\micron$ line is more than 20 times lower than that of H$_{2}$ 1$-$0 S(1) $\lambda$2.122 $\micron$ line. If we consider a higher extinction in $H$-band ($A_{\rm H}$$-$$\Ak$ $\sim$ 2.5), it will correspond to $\sim$ 10 times difference. The channel maps show that the speed of the {\FeII} line is slower than the H$_{2}$ lines. This is different from the cases of other shocked outflows \citep{Pyo2002,Davis2003,Takami2006} or the outer fingers in Orion KL outflow \citep{Bally2015}, where the {\FeII} shows similar or higher velocity than that of H$_{2}$ emission. Also, the {\FeII} ``fingertips'' seen in many outer fingers are not clear in those in the peak 1 region.

Comparison between channel maps of  H$_{2}$ 1$-$0 S(1) $\lambda$2.122 $\micron$ and {\FeII} $\lambda$1.644 $\micron$ lines in Figure \ref{fig:10s1channel} and \ref{fig:Fechannel} gives following results. First, at blueshifted velocities ($|$$\VLSR$$|$ $>$ 65 {\kms}), {\FeII} emission is faint but shows good agreement with outflow features in channel maps of H$_{2}$ 1$-$0 S(1) line. Second, {\FeII} emission does not appear in redshifted velocity channels. This is probably due to higher extinction toward the redshifted fingers (see also Section \ref{sec:ori_Av}). Third, at $|$$\VLSR$$|$ $<$ 65 {\kms}, the two emission lines show different distributions.
In addition, the high-intensity area in the {\FeII} channel map centered at $\VLSR$ $=$ $+$5 {\kms} is consistent with the positions of the several blueshifted and redshifted high-velocity components in H$_{2}$ emission (FID 1, 5, 8, 21, 22, and 29). Since {\FeII} emission usually arises in regions excited by J-type shocks \citep{O'Connell2005,Bally2007}, this positional coincidence indicates the mixture of C- and J-type shocks in the high velocity components. This result is in agreement with that from the H$_{2}$ 2$-$1 / 1$-$0 S(1) line ratio shown in Figure \ref{fig:ratiochannel}.

\section{Discussion}\label{sec:ori_discussion}

\subsection{Velocity gradient in PVDs}
Section \ref{sec:ori_result} shows that the velocity gradients along the 31 finger structures seen at high-velocity are along the outflow direction: $\delta$$v$/$\delta$$l$ $\sim$ 2$-$6 {\kms} arcsec$^{-1}$. There are two possible mechanisms to explain the velocity gradients in the PVDs. The first is that the gradients are part of a global linear velocity variation as a function of distance from the position of driving source \citep{Doi2002,Bally2011}. The second is that each gradient results from local velocities in an expanding bow-shock \citep{Bally2015}. The high-resolution imaging in \citet{Bally2015} showed greater transverse (expansion) velocity at the bow tip than bow tail, and they suggested that shock-heated plasma derives the expansion.

\begin{figure}
\epsscale{.90}
\plotone{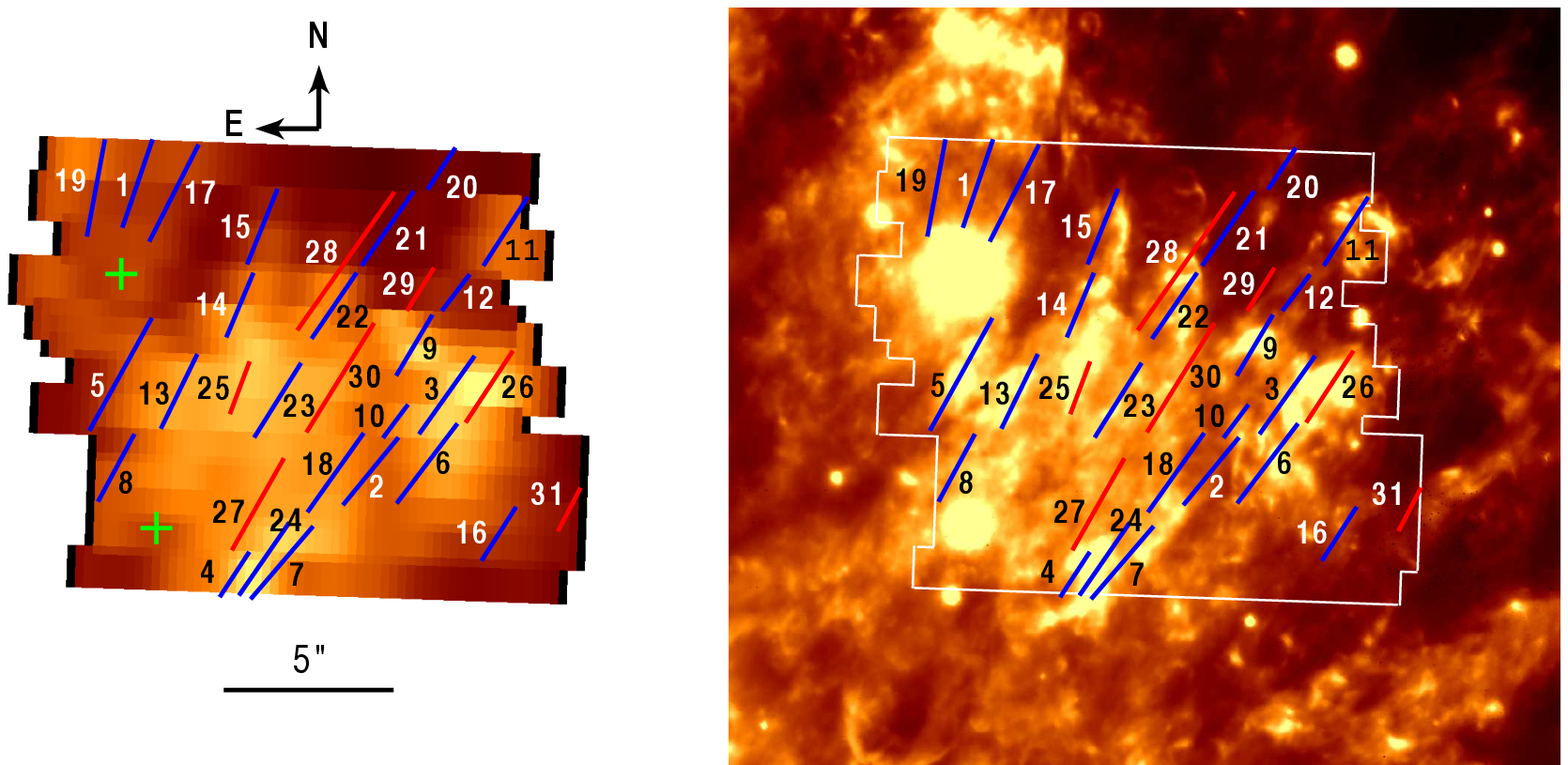}
\caption[(Left) Image of H$_{2}$ 1$-$0 S(1) $\lambda$2.122 $\micron$ emission obtained from observations of 13 consecutive slit positions. (Right) The slit scan area on the high spatial resolution H$_{2}$ 1$-$0 S(1) image of \citet{Bally2015}.]{(Left) Image of H$_{2}$ 1$-$0 S(1) $\lambda$2.122 $\micron$ emission obtained from observations of 13 consecutive slit positions. The image results from integration from ${\VLSR}$ $=$ $-$150 {\kms} to $+$ 150 {\kms}. The two green crosses mark the bright stars in this field, V 2248 Ori and V 1496 Ori \citep{Muench2002}. (Right) The slit scan area as seen in the high spatial resolution H$_{2}$ 1$-$0 S(1) image of \citet{Bally2015}. The white solid line shows the slit-scan area. This image contains continuum, so several of the most prominent features and many less prominent ones are stars. The blue and red lines mark the outflow streams found in our channel maps at blueshifted and redshifted velocity, respectively. The finger identification numbers (FIDs) increase with radial velocity from $-$127 to $+$88 {\kms} (see Table \ref{tbl:fingerlist}). Each line corresponds to the apparent length of the stream identified from the datacube. The intensity scales in both images are linear. \label{fig:arrows}}
\end{figure}

\begin{figure}
\epsscale{0.85}
\plotone{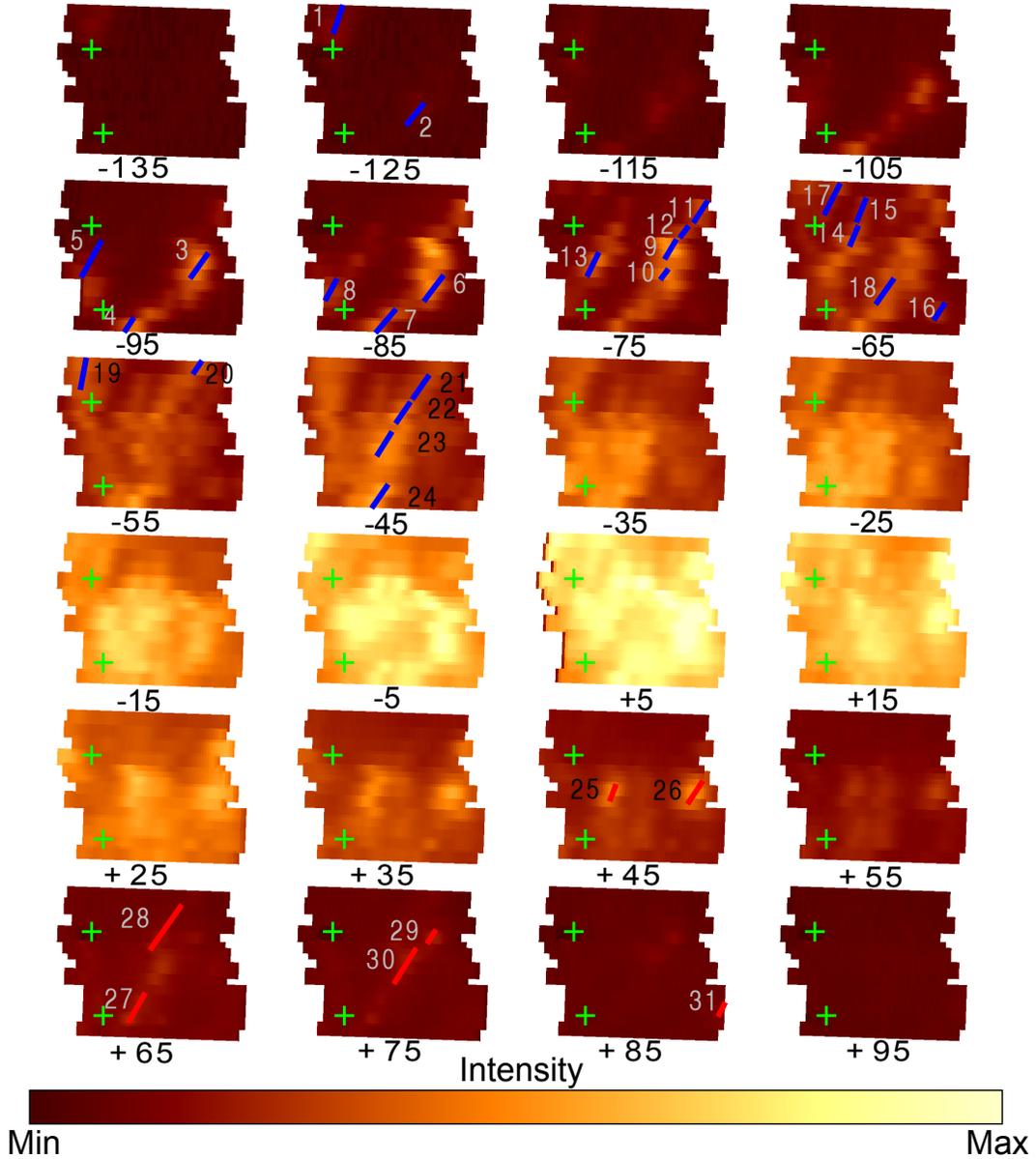}
\caption[Channel maps of the H$_{2}$ 1$-$0 S(1) line. The intensity is integrated over successive 10 {\kms} intervals.]{Channel maps of the H$_{2}$ 1$-$0 S(1) line. The intensity is integrated over successive 10 {\kms} intervals. The central velocity is marked at the bottom of each channel map. The radial velocity increases from top-left to bottom-right. The fingers identified by FIDs are marked with blue and red lines. The intensity is displayed on a square-root scale. \label{fig:10s1channel}}
\end{figure}

\begin{figure}
\epsscale{.60}
\plotone{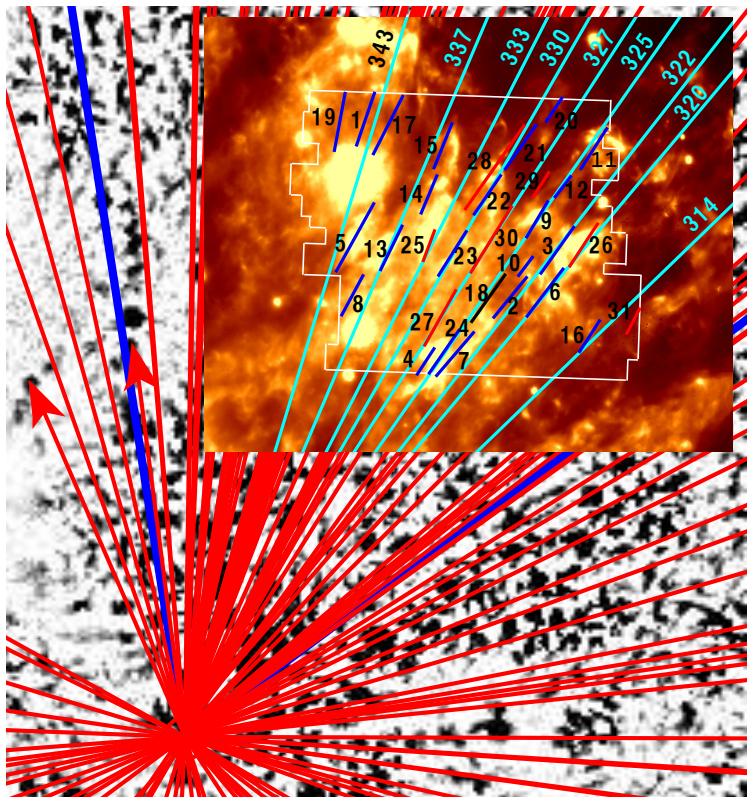}
\caption[Comparison between outflow streams identified in this observation and large scale flow vectors.]{Comparison between outflow streams identified in this observation and large scale flow vectors. The grayscale image with red and thick blue lines is taken from Figure 3 of \citet{Bally2015}. The red and thick blue lines are vectors which connect the ejection center indicated by \citet{Bally2015} and the outermost bow tips in their image. The cyan lines are pseudo-slit positions for the extracted PVDs in Figure \ref{fig:pvd}. These are parallel to the large scale vector. Position angle (P.A.) of cyan lines are marked beside the lines, and is counter-clockwise from the north. The P.A. are 343$\arcdeg$, 337$\arcdeg$, 333$\arcdeg$, 330$\arcdeg$, 327$\arcdeg$, 325$\arcdeg$, 322$\arcdeg$, 320$\arcdeg$, and 314$\arcdeg$, from the left to the right (clockwise direction). \label{fig:ballyvector}}
\end{figure}

\begin{figure}
\epsscale{.9}
\plotone{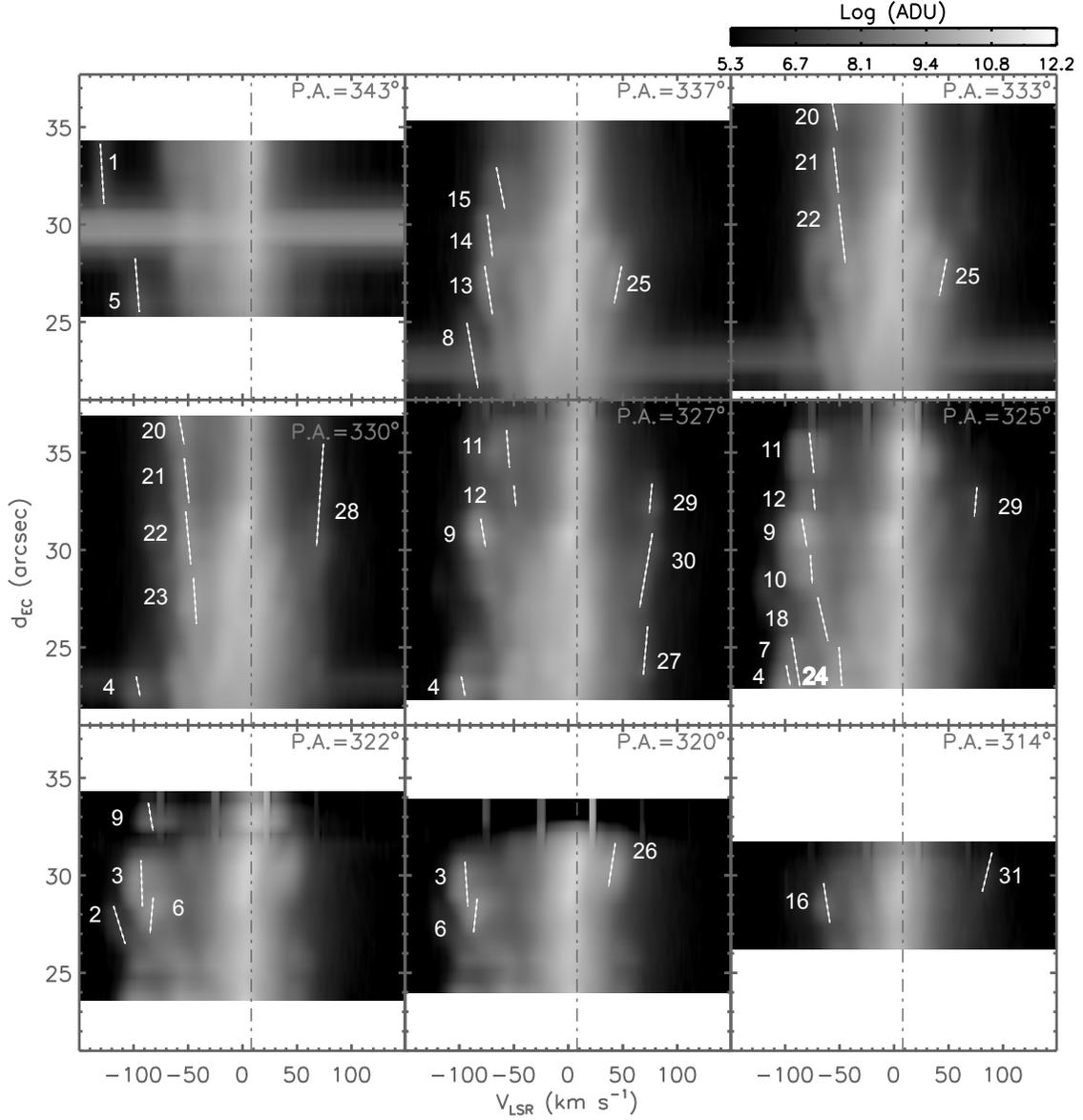}
\caption[PVDs of H$_{2}$ 1$-$0 S(1) line along the cyan lines marked in Figure \ref{fig:ballyvector}.]{PVDs of H$_{2}$ 1$-$0 S(1) line along the cyan lines marked in Figure \ref{fig:ballyvector}. The extraction width of the pseudo-slit is $\sim$1\farcs2. P.A. of the slit is indicated at each panel. The vertical axes indicate the distance as one moves north and west from the ejection center \citep{Bally2015}, along the cyan lines in Figure \ref{fig:ballyvector}.
The white dotted lines and the numbers represent velocity components corresponding to the fingers marked in Figures \ref{fig:arrows} and \ref{fig:10s1channel}. The systemic velocity \citep[${\VLSR}$ $\sim$ $+$8 {\kms},][]{Chrysostomou1997} is indicated with dash-dotted lines. The intensity scale is logarithmic. \label{fig:pvd}}
\end{figure}

\begin{figure}
\epsscale{1.0}
\plotone{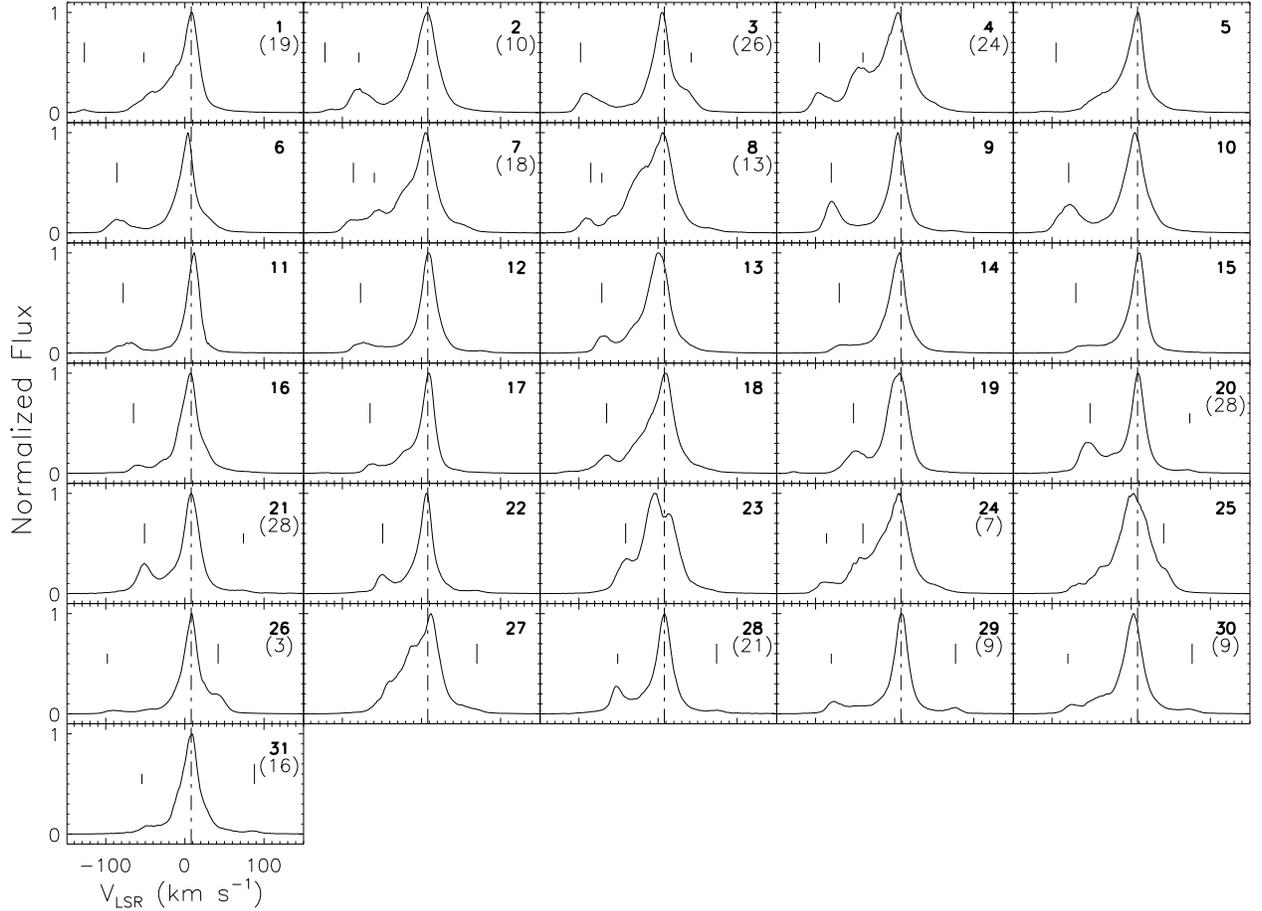}
\caption[Integrated H$_{2}$ 1$-$0 S(1) line profiles at each identified fingers.]{Integrated H$_{2}$ 1$-$0 S(1) line profiles at each identified fingers. The profiles are derived by integrating over $\pm$0\farcs5 of the every fingertip in Figure \ref{fig:arrows} and are normalized to their peak intensity. The FID is marked at upper right corner of each panel. The dominant peak lies near the systemic velocity ($\VLSR$ $\sim$ $+$8 {\kms}) for all regions. All profiles show high velocity peaks that are indicated with long solid vertical lines on the line profiles. At the finger positions where the line profile is contaminated by another finger, the short vertical lines and the numbers within parentheses show the peaks contaminated and their FIDs, respectively.
\label{fig:10s1profile}}
\end{figure}

\begin{figure}
\epsscale{.6}
\plotone{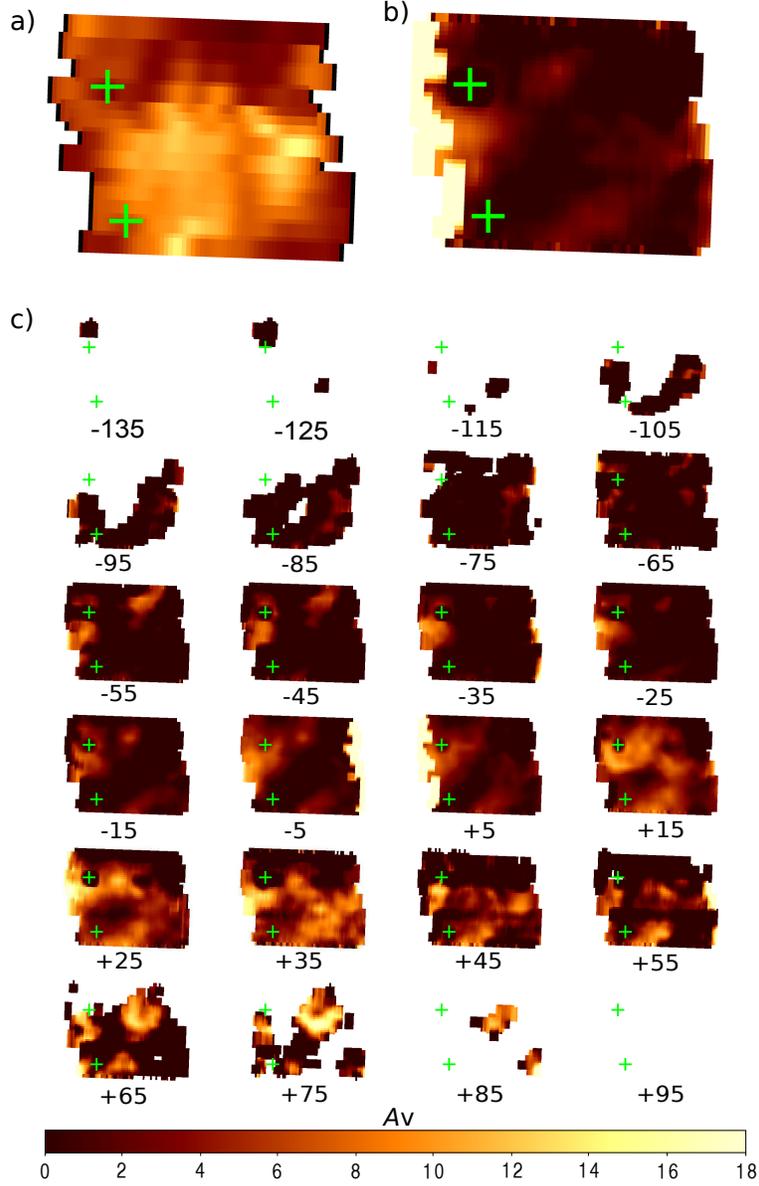}
\caption[(a) H$_{2}$ 1$-$0 S(1) line intensity map, (b) visual extinction ($\Av$) distribution for the H$_{2}$ emission derived from velocity integrated range of $-$150 {\kms} $<$ $\VLSR$ $<$ $+$150 {\kms}, and (c) visual extinctions of channel maps.]{(a) H$_{2}$ 1$-$0 S(1) line intensity map, (b) visual extinction ($\Av$) distribution for the H$_{2}$ emission derived from velocity integrated range of $-$150 {\kms} $<$ $\VLSR$ $<$ $+$150 {\kms}, and (c) visual extinctions of channel maps. We used a median value from three different line ratios of H$_{2}$ v $=$ 1$-$0: Q(3) / S(1), Q(2) / S(0), and Q(4) / S(2). Pixels with low signal-to-noise ratio (S/N $<$3) are excluded in the plot. The extinction maps in (b) and (c) are smoothed with a gaussian mask of 4 $\times$ 4 pixels. The green crosses mark the positions of removed stars. 
\label{fig:Av}}
\end{figure}

\begin{figure}
\epsscale{.85}
\plotone{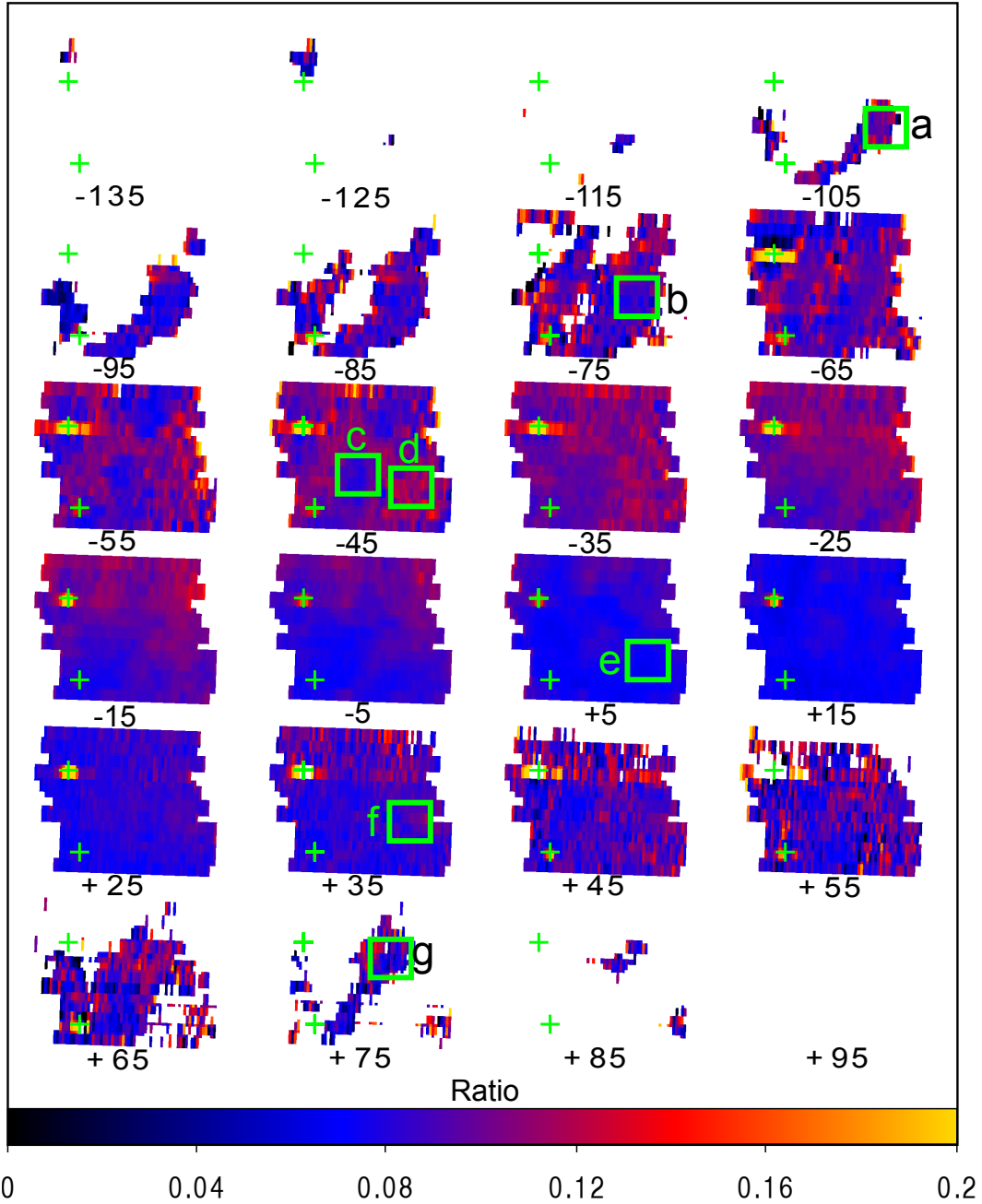}
\caption[Channel maps of H$_{2}$ 2$-$1 S(1) / 1$-$0 S(1) line ratio.]{Channel maps of H$_{2}$ 2$-$1 S(1) / 1$-$0 S(1) line ratio. The ratios show the values expected from shock excitation, where 0.05 and 0.27 are for pure C-shock for J-shock \citep{Smith1995}. In each channel map, reddening is corrected using $\Av$ shown in Figure \ref{fig:Av}. The ratio is close to that from C-type shock models near the systemic velocity ($\VLSR$ = $+$8 {\kms}), but slightly higher than for a pure C-shock.  In both blue and redshifted components, the ratio tends to be larger at higher velocity, indicating a mixture of C- and J-shocks. The green boxes are the integration regions for the population diagrams in Figure \ref{fig:cdr}. Two green crosses mark the positions of removed stars. \label{fig:ratiochannel}}
\end{figure}

\begin{figure}
\epsscale{.9}
\plotone{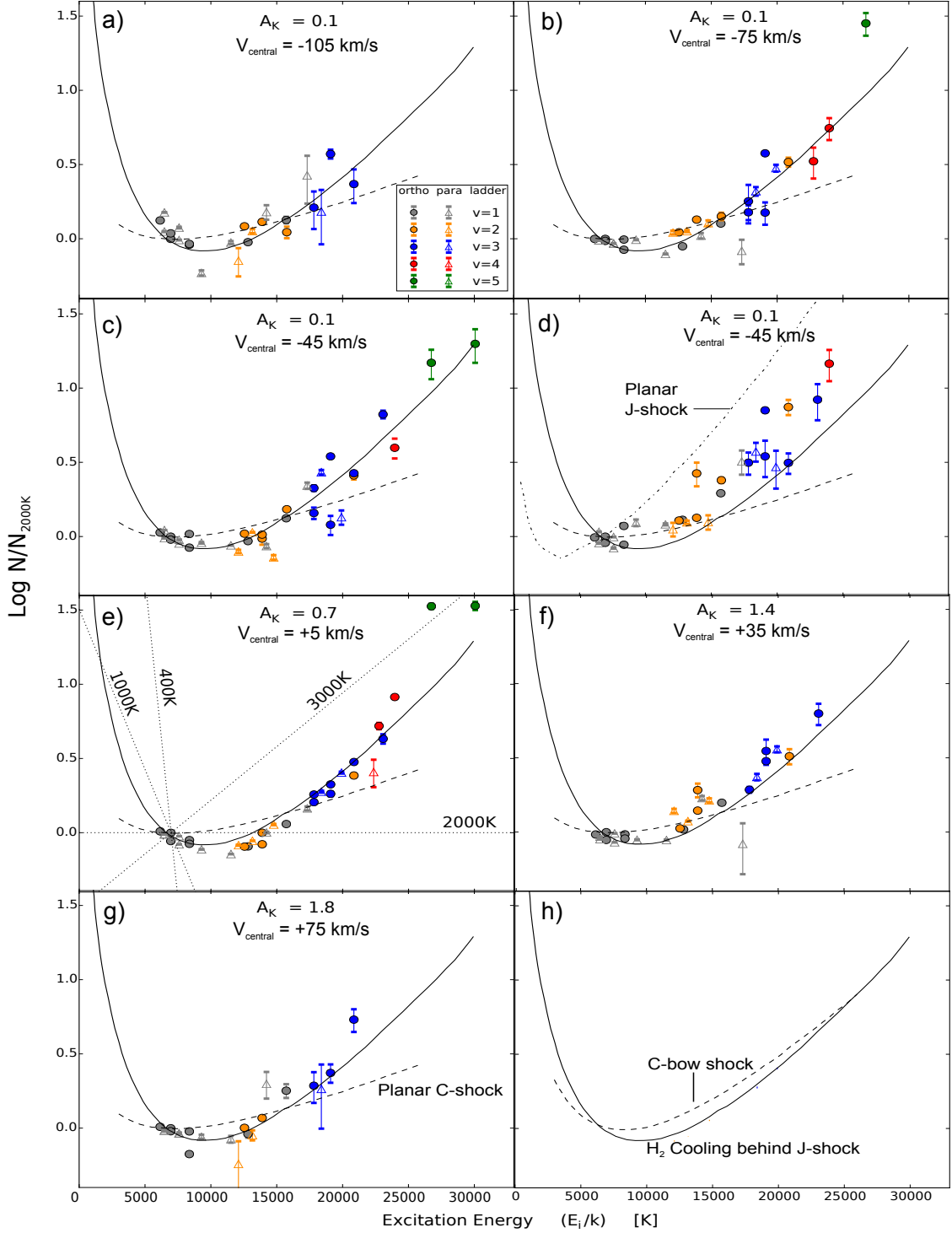}
\caption[Population diagrams at selected position-velocity ranges.]{\scriptsize{Population diagrams at selected position-velocity ranges ($V$$_{\rm central}$ $=$ $-$105, $-$75, $-$45, $+$5, $+$35, and $+$75 {\kms}, $V$$_{\rm width}$ $=$ 10 {\kms}) based on the integrations over the regions indicated with green boxes in Figure \ref{fig:ratiochannel}. The column density values are normalized to H$_{2}$ 1$-$0 S(1) line and are relative to the Boltzmann distribution at 2,000K. In $a)$--$g)$, solid and dashed curves are models of H$_{2}$ cooling zone after J-shock \citep{Brand1988,Burton1989} and C-type planar shock \citep{Smith1991}, respectively. A dash-dotted line in $d)$ is planar J-shock model with conventional cooling \citep{Smith1991,Burton1989}. The dashed line in $h)$ shows C-type bow shock model \citep{Smith1991}. In $e)$, populations at single temperatures of 400, 1,000, 2,000 and 3000 K are shown with dotted straight lines. At $\VLSR$ $=$ $+$5 {\kms}, it shows thermalization at 1,800 (v=1), 2,600 (v=2), 3,200 K (v=3).}\label{fig:cdr}}
\end{figure}

\begin{figure}
\epsscale{.85}
\plotone{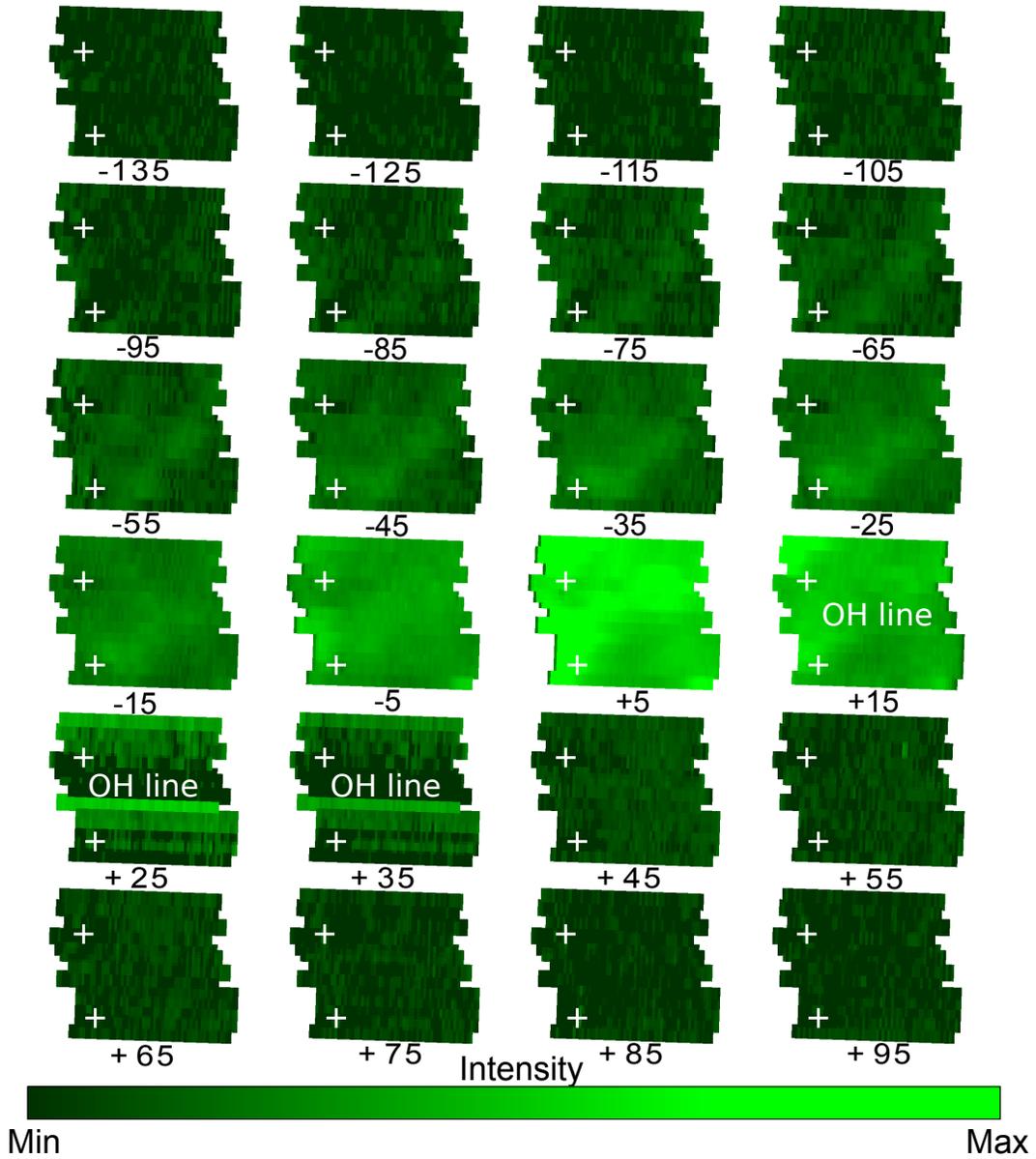}
\caption[Velocity channel maps of {\FeII} 1.644 $\micron$ emission line.]{Velocity channel maps of {\FeII} 1.644 $\micron$ emission line. Velocity ranges of 10 {\kms} $<$ $\VLSR$ $<$ 40 {\kms} are contaminated by residuals after subtraction of strong OH emission line. The intensity is displayed in square-root scale. The white crosses mark the positions of removed stars.
\label{fig:Fechannel}}
\end{figure}

In the introduction, we noted that \citet{Doi2002} and \citet{Bally2011} derived proper motion velocities for HH objects and H$_{2}$ bow fingers that increase linearly with increasing distance from the ejection center.
Our field is  $\sim$ 21$\arcsec$$-$38$\arcsec$ from the ejection center indicated in \citet{Bally2015}, so the proper motion velocity corresponds to 43$-$73 {\kms} based on the fit shown in \citet{Bally2011}. Considering the inclination angles of outflow streams, the average velocity gradient is $\sim$ 4 {\kms} arcsec$^{-1}$. This value agrees with our Doppler velocity-based result of 2$-$6 {\kms} arcsec$^{-1}$, while the expanding bow-shock model would show a significantly larger ($>$20 {\kms}) velocity gradient. Also, the bow-shock expansion might reveal itself as a velocity dispersion rather than as a gradient, but we did not detect a velocity width variation along the outflow direction. We note, however the limitations imposed by our angular resolution. The apparent sizes of the fingers shown in Figure \ref{fig:arrows} are about 2$\arcsec$ in length. The angular resolution along the outflow is larger than that along the slit-scan direction, where the angles between slit and the axes of outflows are 45$\degree$$-$75$\degree$. Further study with higher-angular resolution along the finger axis would helps the interpretation.

The finger groups with FIDs 20, 21, 22, and 23 are aligned in both in space and velocity (Figure \ref{fig:arrows} $\&$ \ref{fig:pvd}). Another group with FIDs 9, 10, 18, and 24 are also aligned. These aligned groups indicate that individual features are related to one large finger as a chain of small bow shocks. The velocity trends of the aligned groups also agree with the global velocity gradient expected from proper motion. The group of FIDs 13, 14, and 15, which is apparently continuous bows in Figure \ref{fig:arrows}, shows a similar velocity gradient but is not continuously connected in PVDs. They might  be moving in somewhat different directions in space.

Chains of bows are a common feature of outflows from low-mass young stars too \citep[e.g., HH 111 and HH 212,][]{Hartigan2001,McCaughrean1994}, and their time variability can produce velocity gradients like those seen in FIDs 13$-$15. However, the outflows from low-mass YSOs show sequential ejections, while  the global velocity pattern in Orion KL is more consistent with the hypothesis of a one-time explosive event  \citep{Bally2005} in this analysis. Also, velocity gradients of different bows which are continuously aligned in PVDs (e.g., FIDs 20$-$23) are not observed in other outflows from low-mass YSOs.
We note the coherence of structures in position-velocity spaces; the lines are narrow ($<$ 30{\kms}) even as $\VLSR$ changes rapidly.
Putting the results above together, the observed PVDs confirm a velocity pattern consistent with explosive dispersal from a single origin.

\subsection{Three-dimensional structure of fingers}\label{sec:ori_3D}
\begin{figure}
\epsscale{1.0}
\plotone{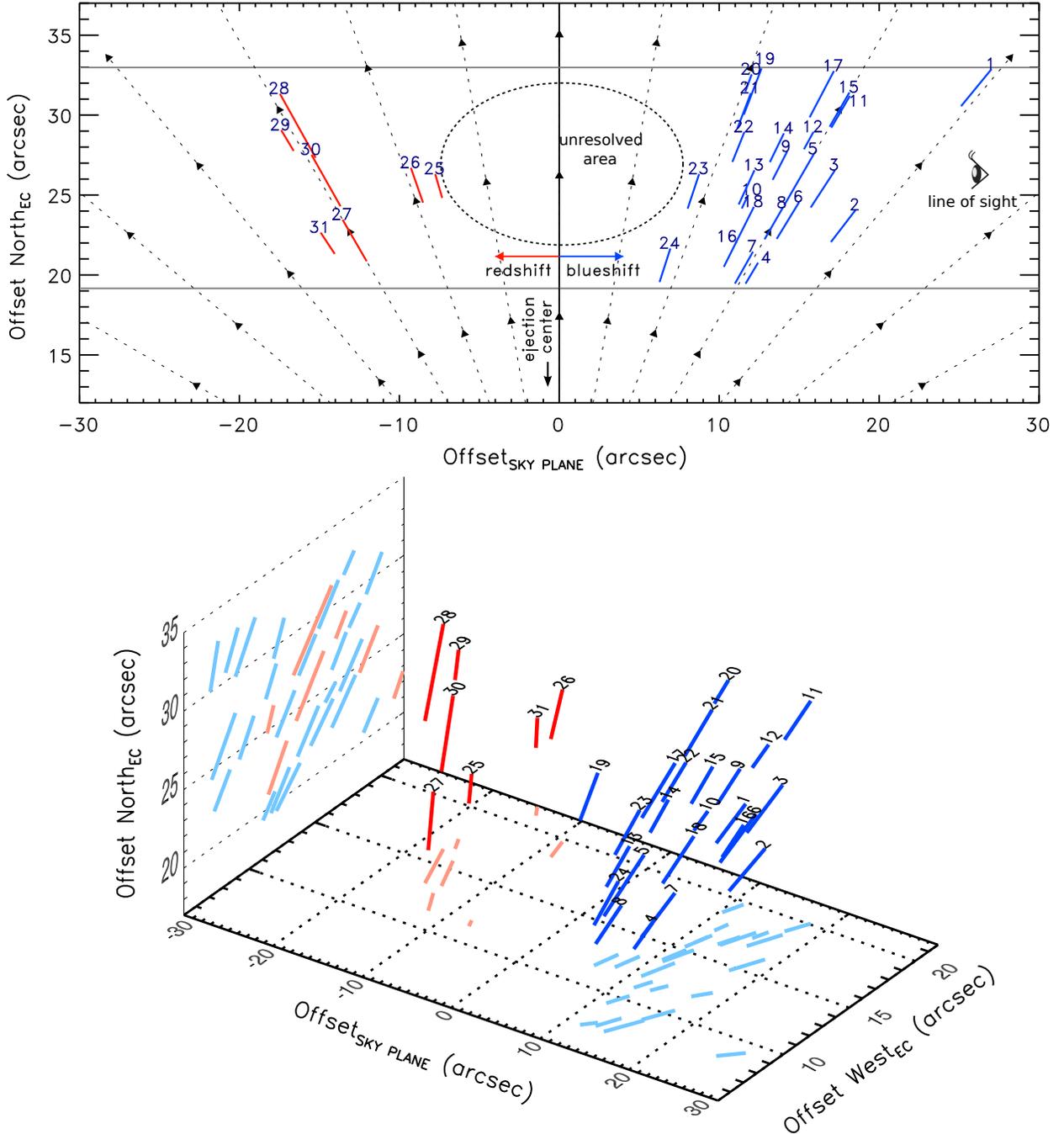}
\caption[(Top) Conceptual two-dimensional drawing of finger distribution along the line of sight. (Bottom) Finger distribution in three-dimensional drawing.]{(Top) Conceptual two-dimensional drawing of finger distribution along the line of sight. We determined finger positions in the space with the estimated outflow inclination angle ($i$) in Table \ref{tbl:fingerlist} and  with an assumption of radial explosion from the ejection center. Y coordinate (Offset North$_{\rm EC}$) is offset distance in north direction from the ejection center indicated in \citet{Bally2015}. Two gray solid lines indicate 19$\arcsec$ $<$ Y $<$ 33$\arcsec$ region covered by slit-scan observation. X coordinate (Offset$_{\rm SKY~PLANE}$) is the offset relative to the plane of the sky, along the line of sight. The lengths of blue and red lines indicate apparent length of stream identified in channel map (Figure \ref{fig:10s1channel}). The interval angle of dotted lines is 10$\degree$. The fingers at 75$\degree$ $< i <$ 90$\degree$ are not resolved, as mentioned in the text. (Bottom) Finger distribution in three-dimension. \label{fig:3d}}
\end{figure}

For the identified fingers, we estimated the speeds of flows using the bow-shock profile to derive the inclination angles.
The double-peak velocity profiles shown in Figure \ref{fig:10s1profile} are well explained by the geometrical bow-shock model \citep[e.g.,][]{Hartigan1987,O'Connell2004}, which shows the emission from the bow tip and wing appear as high- and low-velocity components in the observed profile. \citet{Hartigan1987} indicated that the FWZI of the double-peaked line profile reflects the bullet speed itself. We estimated the FWZI by a multiple-gaussian fitting. Most of the region shows triple-peaks due to an overlap of different fingers (see Figure \ref{fig:10s1profile}). In order to eliminate the contamination by different fingers, we only considered the major peak component and high-velocity component of a target finger in the fitting. The inclination angle ($i$) is derived using the estimated bullet speed and measured peak radial velocity of the high-velocity component. 
We listed measured FWZIs and $i$ angles in Table \ref{tbl:fingerlist}. 
With these inclination angles, we construct a simple 3D map that shows the finger distribution along the line of sight. Figure \ref{fig:3d} shows the constructed map.
We assume that the outflow exploded into all radial directions from the common ejection center. As shown in Table \ref{tbl:fingerlist}, we found the inclination angles of blueshifted and redshifted streams in range of 51$\degree$$-$68$\degree$, with respect to the line of sight. The fingers at 75$\degree$ $<$ $i$ $<$ 90$\degree$ are not resolved because we could not obtain fingers at the velocity range of $-$30 to $+$30 {\kms} (see Section \ref{sec:ori_result}).
We estimate wide-opening angle about 100$\degree$ along the line of sight. This is comparable to the opening angles found in the previous imaging observations \citep[e.g.,][]{Allen1993,Bally2015}.
This confirms that the outflow has a conical shape not only in 2D, but also in 3D. The SiO observations by \citet{Plambeck2009} indicated that the surrounding envelope along a NE-SW axis around radio source I causes the conical shape of the outflow.

\subsection{Internal extinction}\label{sec:ori_Av}
\citet{Rosenthal2000} mentioned the difficulties in determination of infrared extinction curve in OMC-1, due to a mixture of absorbing/emitting gases and complicated outflow distribution. \citet{Scandariato2011} also shows that the extinction of OMC-1 molecular cloud is spatially complicated. Extinctions estimated from atomic and molecular hydrogen lines are very different, $\Ak$ = 0.15 and  0.9, respectively \citep{Rosenthal2000}. The {\ion{H}{1}} traces a different region since they arise in the foreground {\ion{H}{2}} region while the H$_{2}$ lines arise from the deeply embedded cloud behind Orion nebula.

\begin{figure}
\epsscale{.6}
\plotone{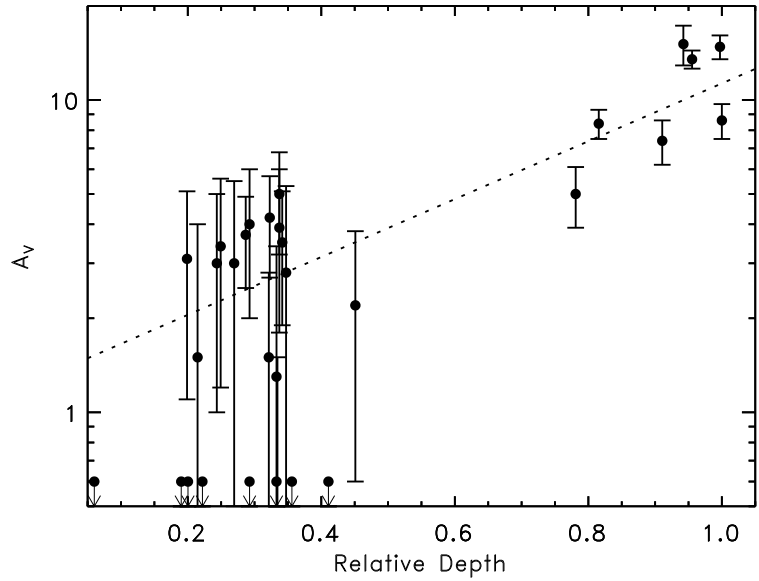}
\caption[Correlation between relative depth along the line of sight and visual extinction $\Av$ for the identified fingers.]{Correlation between relative depth along the line of sight and visual extinction $\Av$ for the identified fingers. The relative depth of each finger is estimated from three-dimensional distribution shown in Figure \ref {fig:3d} and is listed in Table \ref{tbl:fingerlist}. FID 1 and 28 correspond to depth of 0 and 1, respectively. Dotted line represents the fitting line from the linear regression.  FIDs with $\Av$ estimated to be $\sim$ 0 in Table \ref{tbl:fingerlist} are not included in the fitting.
\label{fig:depthvsAv}}
\end{figure}

Section \ref{sec:ori_extinction} showed that the $\Av$ value is low ($\Av$ $=$ 0$-$4.2 mag) and high ($\Av$ $=$ 5.1$-$15.1 mag) at blueshifted and redshifted velocities, respectively. This difference implies an internal extinction between the blueshifted and redshifted fingers.
In connection with 3D distribution map in Figure \ref{fig:3d}, we showed the relation between relative depth and visual extinction in Figure \ref{fig:depthvsAv}. The correlation supports the assumption that the outflows emanate radially from a common center. By adopting a 414 pc as the distance to the OMC-1 cloud \citep{Menten2007}, the estimated maximum and average distances between the blueshifted and redshifted fingers are about 1.8 $\times$ 10$^{4}$ au (0.1 pc) and 1.1 $\times$ 10$^{4}$ au, respectively. The difference between average extinctions ($\Delta$$\Av$) in blueshifted and redshifted fingers  is $\sim$ 8.5 mag. It corresponds to the hydrogen column density $N_{\rm H}$ of $\sim$ 1.6 $\times$ 10$^{22}$ cm$^{-2}$, according to the empirical relation of $N_{\rm H}$/$\Av$ $\approx$ 1.87 $\times$ 10$^{21}$ cm$^{-2}$ mag$^{-1}$ for $R_{\rm V}$ $=$ 3.1 \citep{Bohlin1978,Savage1979,Draine2003}. This is converted into the hydrogen number density ($n$$_{\rm H}$) of $\sim$ 1 $\times$ 10$^{5}$ cm$^{-3}$, and we calculated the total hydrogen mass for the bright ring-shaped area shown in channel maps at $\VLSR$ $=$ $-$5 and $+$5 {\kms} (Figure \ref{fig:10s1channel}). Estimated mass is $\sim$ 0.02 $M$$_{\sun}$ within the cylindrical volume with radius of 6$\arcsec$ (2.5 $\times$ 10$^{3}$ au) and length of 1.1 $\times$ 10$^{4}$ au. The calculated column density $N_{\rm H}$ at each finger location is listed in Table \ref{tbl:fingerlist}. One might expect large grains deep inside the OMC-1 due to the formation of ice mantles and coagulation \citep[e.g.,][]{Pendleton1990}. To consider the larger grain size, we estimate $N_{\rm H}$ with $R_{\rm V}$ $=$ 5. In that case, it gives the smaller $\Delta$$\Av$ of $\sim$ 7.3 and $N_{\rm H}$ of $\sim$ 0.85 $\times$ 10$^{22}$ cm$^{-2}$.

The estimated hydrogen number density of $\sim$ 10$^{5}$ cm$^{-3}$ approaches what we would expect for the dense molecular core \citep[e.g.,][]{Genzel1989}. It is also similar or somewhat smaller than the pre-shock densities used in the model calculations to reproduce the observed line ratios \citep{Chernoff1982,Draine1983,Brand1988,Smith1995}. One might consider the physical effects due to the interactions between the outflows and such a high-density medium, e.g., a deflection in outflows \citep[HH 110 and NGC 1333 IRAS 4A outflow,][]{Reipurth1996,Choi2005}. Simulated jet/cloud collisions by \citet{Raga2002} showed that the high cloud-to-jet density ratio ($\rho_{c}$/$\rho_{j}$) about 100 causes a jet deflection. The large-scale, high spatial resolution images of Orion KL \citep{Bally2015} show that there are no clear bending or distortion of outflows around the peak 1 region. The estimated density at peak 1 is much lower than the densities around the outflow origin, which are 10$^{7}$$-$10$^{8}$ cm$^{-3}$ \citep{Genzel1989}. We suppose that $\rho_{c}$/$\rho_{j}$ is substantially below 100. Furthermore, numerical simulations by \citet{Bally2015} indicated that the bullets should be three orders-of-magnitude denser than the medium to reproduce the bullet shapes shown in their high spatial resolution images of the Orion KL outflows.
One considerable scenario is that some bullets got through the low-density regions in the clumpy medium while the bullets that hit the denser material dispersed and created the shocks at low-velocity, as seen in the diffuse emission at $-$30 {\kms} $<$ ${\VLSR}$ $<$ $+$30 {\kms} in Figure \ref{fig:10s1channel}.

\subsection{Shock excitation}\label{sec:ori_ratio_d}
In Section \ref{sec:ori_ratio}, we showed that the H$_{2}$ line ratios are indicative of the shock excitation and reflect temperatures of  2000$-$3000 K.
We compare the ratios with the empirical fit of J-type H$_{2}$ cooling zone behind a hydrodynamic shock \citep{Brand1988,Burton1989} and a model of planar C-shock \citep{Smith1991}.
In Figure \ref{fig:cdr}$h$, we included a C-type bow shock \citep{Smith1991} that also agrees well with the observed ratios, but requires much stronger magnetic fields (several tens of mG) than observations of Orion KL would indicate \citep{Chrysostomou1994,Tedds1999}.
In all velocity ranges, it is clear that the observed populations are not well-reproduced by a planar C-shock model. They agree well, however, with both J-type H$_{2}$ cooling zone and C-type bow shock models. In fact, the two models do not show a significant difference in the observed excitation energy range. The overall Boltzmann diagrams imply that both C-bow and J-type shock models match the observed population, while the H$_{2}$ 2$-$1 S(1) / 1$-$0 S(1) ratios alone indicate a the mixture of two types of shock. Also, the detection of weak {\FeII} emission supports the possibility of a shock-type mixture. The Far-IR spectral mapping toward peak 1 by \citet{Goicoechea2015} also supports the idea of possible mixture of C- and J-type shocks.

In panel $(d)$, we included one location with fainter emission, i.e., where the outflow streams are not prominent in the channel maps (Figure \ref{fig:10s1channel}). This region shows higher populations of lines from high excitation energies, as expected from the high 2$-$1 S(1) / 1$-$0 S(1) ratio in Figure \ref{fig:ratiochannel}. It indicates the mixture of planar J-shock model with conventional cooling \citep{Smith1991,Burton1997}, indicating fast planar winds. Other fainter, but fast ($|$$\VLSR$$|$ $>$ 40 {\kms}) locations show a similar distribution.

\section{Summary} \label{sec:ori_summary}
We presented the results from high-resolution near-IR spectroscopy toward the Orion KL outflow, and constructed of 3D datacubes for $\sim$ 35 H$_{2}$ ro-vibrational transitions. We summarize the main results as followings:

1. From H$_{2}$ 1$-$0 S(1) datacube, we identified 31 outflow streams that are distinct both kinematically and spatially. We found 24 blueshifted and 7 redshifted streams at $\VLSR$ = $-$130 to $-$40 and $+$45 to $+$90 {\kms}, respectively.

2. PVDs and line profiles indicated that every finger showed multiple velocity peaks at low ($|$$\VLSR$$|$ = 0$-$10 {\kms}) and high velocity ($|$$\VLSR$$|$ = 40$-$130 {\kms}). The low velocity component was always dominant around systemic velocity at $\VLSR$ = $+$8 {\kms}, in agreement with a typical bow shock model.

3. In PVDs, high-velocity components showed a velocity gradient with a decrease of 2$-$6 {\kms} arcsec$^{-1}$ along the direction of finger tip toward the nominal ejection center. This value corresponds to the velocity variation shown in the large-scale proper motion studies, which imply a gradient of $\sim$ 4 {\kms} arcsec$^{-1}$ at inclination angle $i$ $\sim$ 60$\degree$. The combined results further support the scenario of a simultaneous explosive outflow.

4. We constructed a finger distribution map along the line of sight. The inclination angles ($i$) were estimated using the radial velocities and the flow speeds. Fingers were distributed at $i$ $\sim$ 51$-$68$\degree$, while we could not resolve the streams at 75$\degree$ $<$ $i$ $<$ 90$\degree$. It gives a outflow opening angle about 100$\degree$, which confirms a very wide, conical outflow shown in the imaging studies previously.

5. Extinction in each channel map was estimated using H$_{2}$ line ratios. We found the differential extinction depended on the velocity channel ($\Delta$$\Av$ $>$ 10), indicating relatively low and high extinction at blue and redshifted velocities, respectively. This implied that H$_{2}$ bullets of the Orion KL is expanding through a dense medium ($n_{\rm H}$ $\sim$ 10$^{5}$ cm$^{-3}$). The correlation between the relative depths and extinctions again supports the hypothesis of radial explosion from the common origin.

\acknowledgments

This work used the Immersion Grating Infrared Spectrograph (IGRINS) that was developed under a collaboration between the University of Texas at Austin and the Korea Astronomy and Space Science Institute (KASI) with the financial support of the US National Science Foundation under grant AST-1229522, of the University of Texas at Austin, and of the Korean GMT Project of KASI.
This work was partially supported by the National Research
Foundation of Korea (NRF) grant funded by the Korea
Government (MSIP) (No. 2012R1A4A1028713).
This research has made use of the VizieR catalogue access tool, CDS,
 Strasbourg, France. The original description of the VizieR service was
 published in A\&AS 143, 23.


\clearpage



\clearpage

\clearpage

\clearpage

\clearpage

\clearpage

\clearpage

\clearpage

\clearpage

\clearpage

\clearpage

\clearpage

\clearpage

\begin{deluxetable}{lcr}
\tabletypesize{\scriptsize}
\tablecaption{Normalize H$_{2}$ Line Fluxes to the 1$-$0 S(1) line.\label{tbl:H2flux}}
\tablewidth{0pt}
\tablehead{
\colhead{Transition} & \colhead{$\lambda_{\rm vac}$ ($\micron$)} & \colhead{Flux\tablenotemark{a}}}
\startdata
3$-$1 O(5)  & 1.52203 & 0.66   $\pm$ 0.04 \\
2$-$0 O(7)  & 1.54641 & 0.55   $\pm$ 0.04 \\
5$-$3 Q(7) & 1.56263 & 0.18   $\pm$ 0.03 \\
4$-$2 O(4)  & 1.56352 & 0.16   $\pm$ 0.02 \\
5$-$3 O(3) & 1.61354 & 0.24   $\pm$ 0.03 \\
4$-$2 O(5)  & 1.62229 & 0.29   $\pm$ 0.03 \\
3$-$1 O(7)  & 1.64532 & 0.25   $\pm$ 0.02 \\
1$-$0 S(10) & 1.66649 & 0.28   $\pm$ 0.03 \\
1$-$0 S(9)  & 1.68772 & 2.92   $\pm$ 0.03 \\
1$-$0 S(8)  & 1.71466 & 2.28   $\pm$ 0.05 \\
1$-$0 S(7)  & 1.74803 & 12.69  $\pm$ 0.07 \\
1$-$0 S(6)  & 1.78795 & 7.48   $\pm$ 0.04 \\
2$-$1 S(9)  & 1.79041 & 0.30   $\pm$ 0.03 \\
2$-$1 S(5)  & 1.94487 & 8.78   $\pm$ 0.38 \\
3$-$2 S(7)  & 1.96922 & 87.30  $\pm$ 0.31 \\
1$-$0 S(3)  & 1.95756 & 0.38   $\pm$ 0.06 \\
2$-$1 S(4)  & 2.00407 & 2.46   $\pm$ 0.05 \\
1$-$0 S(2)  & 2.03376 & 35.27  $\pm$ 0.06 \\
3$-$2 S(5)  & 2.06556 & 0.86   $\pm$ 0.02 \\
2$-$1 S(3)  & 2.07351 & 8.82   $\pm$ 0.03 \\
1$-$0 S(1)  & 2.12183 & 100.00 $\pm$ 0.09 \\
3$-$2 S(4)  & 2.12797 & 0.39   $\pm$ 0.01 \\
2$-$1 S(2)  & 2.15423 & 2.89   $\pm$ 0.02 \\
3$-$2 S(3)  & 2.20140 & 1.23   $\pm$ 0.02 \\
1$-$0 S(0)  & 2.22330 & 21.67  $\pm$ 0.04 \\
2$-$1 S(1)  & 2.24772 & 8.13   $\pm$ 0.02 \\
3$-$2 S(2)  & 2.28703 & 0.40   $\pm$ 0.01 \\
4$-$3 S(3)  & 2.34448 & 0.28   $\pm$ 0.01 \\
2$-$1 S(0)  & 2.35563 & 1.33   $\pm$ 0.02 \\
3$-$2 S(1)  & 2.38645 & 0.88   $\pm$ 0.02 \\
1$-$0 Q(1)  & 2.40659 & 72.83  $\pm$ 0.08 \\
1$-$0 Q(2)  & 2.41344 & 22.91  $\pm$ 0.05 \\
1$-$0 Q(3)  & 2.42373 & 62.94  $\pm$ 0.10 \\
1$-$0 Q(4)  & 2.43749 & 17.89  $\pm$ 0.04 \\
1$-$0 Q(5)  & 2.45475 & 42.01  $\pm$ 0.13 \\
1$-$0 Q(6)  & 2.47554 & 8.85   $\pm$ 0.07 \\
\enddata
\tablenotetext{a}{Reddening corrected ($A_{\rm V}$ = 6 mag), normalized flux. 1$-$0 S(1) line flux is set to 100. Flux is integrated at $\pm$ 1$\arcsec$ of brightest peak at FID 26 (see Figure \ref{fig:arrows}).}
\end{deluxetable}

\begin{deluxetable}{ccccrccccc}
\tabletypesize{\scriptsize}
\tablecaption{Identified finger list.\label{tbl:fingerlist}}
\tablewidth{0pt}
\tablehead{
\colhead{FID\tablenotemark{a}} &\colhead{$\alpha$, $\delta$$\tablenotemark{b}$} &\colhead{PA\tablenotemark{c}} &\colhead{length} & \colhead{$v_{\rm peak}$\tablenotemark{d}} & \colhead{FWZI } & \colhead{$i$\tablenotemark{e}}& \colhead{$\Av$} & \colhead{$N_{\rm H}$} & \colhead{relative}\\
 &\colhead{(J2000)}&\colhead{($\degree$)} &\colhead{($\arcsec$)}& \colhead{({\kms})} & \colhead{({\kms})} & \colhead{($\degree$)}& \colhead{(mag)} & \colhead{(10$^{22}$ cm$^{-2}$)} &depth

}
\startdata
1    &5:35:13.809, -5:21:58.35 &341&2.7& $-$127 (14)    & 200  $\pm$ 41        & 51 $\pm$ 8      & \textless0        & -    & 0.00           \\
2   &5:35:13.374, -5:22:06.52 &321&2.6& $-$122 (19)   & 200  $\pm$ 46        & 52 $\pm$ 8      & \textless0        & -    & 0.19           \\
3   &5:35:13.224, -5:22:04.43 &325&2.8& $-$98 (17)    & 180  $\pm$ 55        & 57 $\pm$ 8      & 1.5 $\pm$ 2.5     & 0.3  & 0.21           \\
4   &5:35:13.617, -5:22:09.23 &327&1.6& $-$95 (21)    & 185  $\pm$ 46        & 59 $\pm$ 7      & 4.2 $\pm$ 1.5     & 0.8  & 0.32           \\
5   &5:35:13.874, -5:22:04.33 &331&3.8& $-$95 (33)    & 190  $\pm$ 45        & 60 $\pm$ 6      & 3.0 $\pm$ 2.1     & 0.5  & 0.24           \\
6   &5:35:13.268, -5:22:06.47&323&3.0& $-$86 (23)    & 165  $\pm$ 47        & 59 $\pm$ 8      & 3.1 $\pm$ 2.5     & 0.6  & 0.27           \\
7  &5:35:13.556, -5:22:09.29&319&2.8& $-$86 (24)    & 175  $\pm$ 51        & 61 $\pm$ 7      & 0.4 $\pm$ 1.1     & 0.1  & 0.33           \\
8  &5:35:13.857, -5:22:06.42&332& 2.3& $-$85 (20)    & 170  $\pm$ 45        & 60 $\pm$ 7      & 4.0 $\pm$ 2.2     & 0.7  & 0.29           \\
9  &5:35:13.269, -5:22:02.71&329& 2.1& $-$80 (20)    & 175  $\pm$ 45        & 63 $\pm$ 6      & 3.7 $\pm$ 1.2     & 0.7  & 0.29           \\
10 &5:35:13.295, -5:22:04.54&323& 1.2& $-$79 (26)    & 185  $\pm$ 51        & 65 $\pm$ 6      & 3.5 $\pm$ 1.6     & 0.6  & 0.34           \\
11  &5:35:13.097, -5:21:59.47&327&2.4& $-$78 (25)    & 155  $\pm$ 36        & 60 $\pm$ 6      & 3.1 $\pm$ 2       & 0.6  & 0.20           \\
12  &5:35:13.179, -5:22:00.80&323&1.3& $-$77 (29)    & 160  $\pm$ 45        & 61 $\pm$ 7      & 3.4 $\pm$ 2.2     & 0.6  & 0.25           \\
13  &5:35:13.733, -5:22:04.27&334&2.4& $-$71 (20)    & 170  $\pm$ 43        & 65 $\pm$ 5      & 1.3 $\pm$ 2.1     & 0.2  & 0.33           \\
14 &5:35:13.604, -5:22:01.58&337&2.1& $-$70 (33)    & 160  $\pm$ 36        & 64 $\pm$ 5      & \textless0        & -    & 0.29           \\
15  &5:35:13.563, -5:21:59.43&338&2.4& $-$70 (33)    & 140  $\pm$ 36        & 60 $\pm$ 7      & \textless0        & -    & 0.20           \\
16  &5:35:13.101, -5:22:08.16&327&1.9& $-$65 (19)    & 145  $\pm$ 36        & 63 $\pm$ 6      & \textless0        & -    & 0.36           \\
17  &5:35:13.756, -5:21:58.76&333&3.2& $-$65 (22)    & 140  $\pm$ 39        & 62 $\pm$ 6      & \textless0        & -    & 0.22           \\
18  &5:35:13.448, -5:22:06.73&325&2.9& $-$65 (26)    & 145  $\pm$ 54        & 63 $\pm$ 8      & \textless0        & -    & 0.33           \\
19  &5:35:13.879, -5:21:58.61&349&2.9& $-$52 (27)    & 145  $\pm$ 51        & 69 $\pm$ 6      & 1.5 $\pm$1.3     & 0.3  & 0.32           \\
20  &5:35:13.207, -5:21:57.22&326&1.5& $-$52 (30)    & 150  $\pm$ 51        & 70 $\pm$ 5      & 3.9 $\pm$ 2.1     & 0.7  & 0.34           \\
21  &5:35:13.337, -5:21:59.47&325&2.7& $-$51 (21)    & 143  $\pm$ 50        & 69 $\pm$ 6      & 5.0 $\pm$ 1.8     & 0.9  & 0.34           \\
22  &5:35:13.436, -5:22:01.62&326&2.4& $-$49 (23)    & 132  $\pm$ 54        & 68 $\pm$ 7      & 2.8 $\pm$ 2.5     & 0.5  & 0.35           \\
23  &5:35:13.549, -5:22:04.53&328&2.6& $-$41 (33)    & 130  $\pm$ 60        & 72 $\pm$ 6      & \textless0        & -    & 0.41           \\
24  &5:35:13.579, -5:22:09.18&325&2.6& $-$40 (32)    & 131  $\pm$ 57        & 72 $\pm$ 5      & 2.2 $\pm$ 1.6     & 0.4  & 0.45           \\
25  &5:35:13.597, -5:22:03.84&339&1.7& 41 (19)     & 145  $\pm$ 54        & 74 $\pm$ 5      & 5.1 $\pm$ 1.1     & 0.9  & 0.78           \\
26  &5:35:13.132, -5:22:04.09&327&2.5& 42 (24)     & 128  $\pm$ 45        & 71 $\pm$ 5      & 8.4 $\pm$ 0.9     & 1.5  & 0.82           \\
27  &5:35:13.593, -5:22:07.85&330&3.1& 70 (19)     & 140  $\pm$ 52        & 60 $\pm$ 9      & 7.4 $\pm$ 1.2     & 1.3  & 0.91           \\
28  &5:35:13.464, -5:22:01.37&325&5.0& 74 (23)     & 152  $\pm$ 56        & 61 $\pm$ 8      & 8.6 $\pm$ 1.1     & 1.5  & 1.00           \\
29  &5:35:13.248, -5:22:00.80&328&1.5& 77 (19)     & 150  $\pm$ 54        & 59 $\pm$ 9      & 14.8 $\pm$ 1.3     & 2.7  & 1.00           \\
30  &5:35:13.446, -5:22:04.38&328&3.8& 77 (26)     & 157  $\pm$ 47        & 61 $\pm$ 7      & 13.5 $\pm$ 0.9     & 2.4  & 0.96           \\
31  &5:35:12.950, -5:22:07.29&333&1.4& 88 (20)      & 160  $\pm$ 54        & 57 $\pm$ 9      & 15.1 $\pm$ 2.2     & 2.7  & 0.94          \\
\enddata
\tablenotetext{a}{Finger identification number.}
\tablenotetext{b}{Coordinate at the start point of each finger, i.e., the southeast end of lines in Figure \ref{fig:arrows} and \ref{fig:10s1channel}. The uncertainty is $\pm$0\farcs5.}
\tablenotetext{c}{Position angle of finger in counter clockwise from the north.}
\tablenotetext{d}{Peak radial velocity of high-velocity component in line profile. The value within parenthesis is FWHM of each component.}
\tablenotetext{e}{Inclination angle of outflow with respect to the line of sight.}
\end{deluxetable}


\begin{thebibliography}{}


\bibitem[Allen 
\& Burton(1993)]{Allen1993} Allen, D.~A., \& Burton, M.~G.\ 1993, \nat, 363, 54 

\bibitem[Axon 
\& Taylor(1984)]{Axon1984} Axon, D.~J., \& Taylor, K.\ 1984, \mnras, 207, 241 

\bibitem[Bally et al.(2011)]{Bally2011} Bally, J., Cunningham, 
N.~J., Moeckel, N., et al.\ 2011, \apj, 727, 113 

\bibitem[Bally et 
al.(2015)]{Bally2015} Bally, J., Ginsburg, A., Silvia, D., \& Youngblood, A.\ 2015, \aap, 579, A130 

\bibitem[Bally et al.(2007)]{Bally2007} Bally, J., Reipurth, B., 
\& Davis, C.~J.\ 2007, Protostars and Planets V, 215 


\bibitem[Bally 
\& Zinnecker(2005)]{Bally2005} Bally, J., \& Zinnecker, H.\ 2005, \aj, 129, 2281 

\bibitem[Beckwith et al.(1983)]{Beckwith1983} Beckwith, S., Evans, 
N.~J., II, Gatley, I., Gull, G., \& Russell, R.~W.\ 1983, \apj, 264, 152 

\bibitem[Beckwith et al.(1978)]{Beckwith1978} Beckwith, S., Persson, 
S.~E., Neugebauer, G., \& Becklin, E.~E.\ 1978, \apj, 223, 464 

\bibitem[Becklin 
\& Neugebauer(1967)]{Becklin1967} Becklin, E.~E., \& Neugebauer, G.\ 1967, \apj, 147, 799 

\bibitem[Black 
\& van Dishoeck(1987)]{Black1987} Black, J.~H., \& van Dishoeck, E.~F.\ 1987, \apj, 322, 412 

\bibitem[Black 
\& Dalgarno(1976)]{Black1976} Black, J.~H., \& Dalgarno, A.\ 1976, \apj, 203, 132 

\bibitem[Bohlin et al.(1978)]{Bohlin1978} Bohlin, R.~C., Savage, 
B.~D., \& Drake, J.~F.\ 1978, \apj, 224, 132

\bibitem[Brand et al.(1988)]{Brand1988} Brand, P.~W.~J.~L., 
Moorhouse, A., Burton, M.~G., et al.\ 1988, \apjl, 334, L103

\bibitem[Burton et al.(1989)]{Burton1989} Burton, M., Brand, P., 
Moorhouse, A., 
\& Geballe, T.\ 1989, Infrared Spectroscopy in Astronomy, 290, 281

\bibitem[Burton \& Haas(1997)]{Burton1997} Burton, M.~G., \& Haas, M.~R.\ 1997, \aap, 327, 309 

\bibitem[Chatterjee 
\& Tan(2012)]{Chatterjee2012} Chatterjee, S., \& Tan, J.~C.\ 2012, \apj, 754, 152 

\bibitem[Chernoff et al.(1982)]{Chernoff1982} Chernoff, D.~F., McKee, C.~F., \& Hollenbach, D.~J.\ 1982, \apjl, 259, L97 

\bibitem[Choi(2005)]{Choi2005} Choi, M.\ 2005, \apj, 630, 976 

\bibitem[Chrysostomou et al.(1997)]{Chrysostomou1997} Chrysostomou, A., 
Burton, M.~G., Axon, D.~J., et al.\ 1997, \mnras, 289, 605

\bibitem[Chrysostomou et al.(1994)]{Chrysostomou1994} Chrysostomou, A., Hough, J.~H., Burton, M.~G., \& Tamura, M.\ 1994, \mnras, 268, 325 

\bibitem[Cohen et al.(2006)]{Cohen2006} Cohen, R.~J., Gasiprong, 
N., Meaburn, J., \& Graham, M.~F.\ 2006, \mnras, 367, 541 

\bibitem[Davis et 
al.(2003)]{Davis2003} Davis, C.~J., Whelan, E., Ray, T.~P., \& Chrysostomou, A.\ 2003, \aap, 397, 693 

\bibitem[Doi et al.(2002)]{Doi2002} Doi, T., O'Dell, C.~R., 
\& Hartigan, P.\ 2002, \aj, 124, 445 



\bibitem[Draine(2003)]{Draine2003} Draine, B.~T.\ 2003, \araa, 41, 241

\bibitem[Draine et al.(1983)]{Draine1983} Draine, B.~T., Roberge, W.~G., \& Dalgarno, A.\ 1983, \apj, 264, 485 

\bibitem[Everett et al.(1995)]{Everett1995} Everett, M.~E., Depoy, 
D.~L., \& Pogge, R.~W.\ 1995, \aj, 110, 1295 

\bibitem[Genzel et al.(1981)]{Genzel1981} Genzel, R., Reid, M.~J., 
Moran, J.~M., \& Downes, D.\ 1981, \apj, 244, 884 

\bibitem[Genzel \& Stutzki(1989)]{Genzel1989} Genzel, R., \& Stutzki, J.\ 1989, \araa, 27, 41 

\bibitem[Goicoechea et al.(2015)]{Goicoechea2015} Goicoechea, J.~R., Chavarr{\'{\i}}a, L., Cernicharo, J., et al.\ 2015, \apj, 799, 102 


\bibitem[G{\'o}mez et al.(2005)]{Gomez2005} G{\'o}mez, L., Rodr{\'{\i}}guez, L.~F., Loinard, L., et al.\ 2005, \apj, 635, 1166 

\bibitem[G{\'o}mez et al.(2008)]{Gomez2008} G{\'o}mez, L., 
Rodr{\'{\i}}guez, L.~F., Loinard, L., et al.\ 2008, \apj, 685, 333-343 

\bibitem[Hartigan et al.(1987)]{Hartigan1987} Hartigan, P., Raymond, 
J., \& Hartmann, L.\ 1987, \apj, 316, 323 

\bibitem[Hartigan et al.(2001)]{Hartigan2001} Hartigan, P., Morse, J.~A., Reipurth, B., Heathcote, S., \& Bally, J.\ 2001, \apjl, 559, L157 

Oh, J.~S., et al.\ 2014, \procspie, 9154, 91541X 

\bibitem[Kleinmann 
\& Low(1967)]{Kleinmann1967} Kleinmann, D.~E., \& Low, F.~J.\ 1967, \apjl, 149, L1 


\bibitem[Knapp et al.(1981)]{Knapp1981} Knapp, G.~R., Phillips, 
T.~G., Redman, R.~O., \& Huggins, P.~J.\ 1981, \apj, 250, 175 


\bibitem[Kwan 
\& Scoville(1976)]{Kwan1976} Kwan, J., \& Scoville, N.\ 1976, \apjl, 210, L39 

\bibitem[McCaughrean et al.(1994)]{McCaughrean1994} McCaughrean, M.~J., Rayner, J.~T., \& Zinnecker, H.\ 1994, \apjl, 436, L189 

\bibitem[Matthews et al.(2010)]{Matthews2010} Matthews, L.~D., 
Greenhill, L.~J., Goddi, C., et al.\ 2010, \apj, 708, 80 

\bibitem[Menten \& Reid(1995)]{Menten1995} Menten, K.~M., \& Reid, M.~J.\ 1995, \apjl, 445, L157 

\bibitem[Menten et 
al.(2007)]{Menten2007} Menten, K.~M., Reid, M.~J., Forbrich, J., \& Brunthaler, A.\ 2007, \aap, 474, 515 

\bibitem[Muench et al.(2002)]{Muench2002} Muench, A.~A., Lada, E.~A., Lada, C.~J., \& Alves, J.\ 2002, \apj, 573, 366 

\bibitem[Nisini et 
al.(2002)]{Nisini2002} Nisini, B., Caratti o Garatti, A., Giannini, T., \& Lorenzetti, D.\ 2002, \aap, 393, 1035 

\bibitem[O'Connell et 
al.(2004)]{O'Connell2004} O'Connell, B., Smith, M.~D., Davis, C.~J., et al.\ 2004, \aap, 419, 975 

\bibitem[O'Connell et al.(2005)]{O'Connell2005} O'Connell, B., Smith, M.~D., Froebrich, D., Davis, C.~J., \& Eisl{\"o}ffel, J.\ 2005, \aap, 431, 223 

\bibitem[Oh et al.(2016)]{Oh2016} Oh, H., Pyo, T.-S., Yuk, I.-S., et al.\ 2016, \apj, 817, 148 

S.-M., et al.\ 2014, Journal of Astronomy and Space Sciences, 31, 177

\bibitem[Pak et al.(1998)]{Pak1998} Pak, S., Jaffe, D.~T., van Dishoeck, E.~F., Johansson, L.~E.~B., \& Booth, R.~S.\ 1998, \apj, 498, 735 

\bibitem[Park et al.(2014)]{Park2014} Park, C., Jaffe, D.~T., 
Yuk, I.-S., et al.\ 2014, \procspie, 9147, 91471D 

\bibitem[Pendleton et al.(1990)]{Pendleton1990} Pendleton, Y.~J., Tielens, A.~G.~G.~M., \& Werner, M.~W.\ 1990, \apj, 349, 107

\bibitem[Peng et 
al.(2012)]{Peng2012} Peng, T.-C., Despois, D., Brouillet, N., Parise, B., \& Baudry, A.\ 2012, \aap, 543, A152 

\bibitem[Plambeck et al.(2009)]{Plambeck2009} Plambeck, R.~L., Wright, M.~C.~H., Friedel, D.~N., et al.\ 2009, \apjl, 704, L25 

\bibitem[Plambeck et al.(1982)]{Plambeck1982} Plambeck, R.~L., 
Wright, M.~C.~H., Welch, W.~J., et al.\ 1982, \apj, 259, 617 

\bibitem[Pyo et al.(2002)]{Pyo2002} Pyo, T.-S., Hayashi, M., 
Kobayashi, N., et al.\ 2002, \apj, 570, 724

\bibitem[Raga et al.(2002)]{Raga2002} Raga, A.~C., de Gouveia Dal Pino, E.~M., Noriega-Crespo, A., Mininni, P.~D., \& Vel{\'a}zquez, P.~F.\ 2002, \aap, 392, 267 

\bibitem[Reipurth et al.(1996)]{Reipurth1996} Reipurth, B., Raga, A.~C., \& Heathcote, S.\ 1996, \aap, 311, 989 

\bibitem[Rieke 
\& Lebofsky(1985)]{Rieke1985} Rieke, G.~H., \& Lebofsky, M.~J.\ 1985, \apj, 288, 618 

\bibitem[Rosenthal et 
al.(2000)]{Rosenthal2000} Rosenthal, D., Bertoldi, F., \& Drapatz, S.\ 2000, \aap, 356, 705

\bibitem[Salas et al.(1999)]{Salas1999} Salas, L., Rosado, M., 
Cruz-Gonz{\'a}lez, I., et al.\ 1999, \apj, 511, 822 

\bibitem[Savage 
\& Mathis(1979)]{Savage1979} Savage, B.~D., \& Mathis, J.~S.\ 1979, \araa, 17, 73 

\bibitem[Scandariato et 
al.(2011)]{Scandariato2011} Scandariato, G., Robberto, M., Pagano, I., \& Hillenbrand, L.~A.\ 2011, \aap, 533, A38

\bibitem[Smith(1991)]{Smith1991} Smith, M.~D.\ 1991, \mnras, 253, 
175 

\bibitem[Smith(1995)]{Smith1995} Smith, M.~D.\ 1995, \aap, 296, 789 

\bibitem[Sugai et al.(1994)]{Sugai1994} Sugai, H., Usuda, T., Kataza, H., et al.\ 1994, \apj, 420, 746 

\bibitem[Takami et al.(2006)]{Takami2006} Takami, M., 
Chrysostomou, A., Ray, T.~P., et al.\ 2006, \apj, 641, 357 

\bibitem[Tan(2004)]{Tan2004} Tan, J.~C.\ 2004, \apjl, 607, L47 

\bibitem[Taylor et al.(1984)]{Taylor1984} Taylor, K.~N.~R., 
Storey, J.~W.~V., Sandell, G., Williams, P.~M., 
\& Zealey, W.~J.\ 1984, \nat, 311, 236 

\bibitem[Tedds et al.(1999)]{Tedds1999} Tedds, J.~A., Brand, P.~W.~J.~L., \& Burton, M.~G.\ 1999, \mnras, 307, 337 

\bibitem[Turner et al.(1977)]{Turner1977} Turner, J., 
Kirby-Docken, K., \& Dalgarno, A.\ 1977, \apjs, 35, 281 

\bibitem[Usuda et al.(1996)]{Usuda1996} Usuda, T., Sugai, H., 
Kawabata, H., et al.\ 1996, \apj, 464, 818 

\bibitem[Welch et al.(1981)]{Welch1981} Welch, W.~J., Wright, 
M.~C.~H., Plambeck, R.~L., Bieging, J.~H., 
\& Baud, B.\ 1981, \apjl, 245, L87 


\bibitem[Wu et al.(2014)]{Wu2014} Wu, Y., Liu, T., 
\& Qin, S.-L.\ 2014, \apj, 791, 123 

\bibitem[Youngblood et al.(2016)]{Youngblood2016} Youngblood, A., Ginsburg, A., \& Bally, J.\ 2016, \aj, 151, 173 

\bibitem[Yuk et al.(2010)]{Yuk2010} Yuk, I.-S., Jaffe, D.~T., 
Barnes, S., et al.\ 2010, \procspie, 7735, 77351M 





\end{thebibliography}
\end{document}